\documentclass[showpacs,amsmath,amssymb,twocolumn,prl,superscriptaddress,longbibliography]{revtex4-1}
\usepackage{amssymb}
\usepackage[dvips]{graphicx}
\usepackage{enumerate}
\usepackage{epsfig}
\usepackage{subfigure}
\usepackage{xcolor}
\usepackage[T1]{fontenc}
\usepackage{fullpage}
\usepackage{amsthm,amsfonts,amssymb,amscd,mathrsfs,xspace,framed}
\usepackage{amsmath}
\usepackage{color}
\usepackage{setspace}
\usepackage{url}
\usepackage{wrapfig}
\usepackage{enumitem}

\begin{document}


\title{Scalable photonic-phonoinc integrated circuitry for reconfigurable signal processing}


\author{Liang Zhang}
\affiliation{Chandra Department of Electrical and Computer Engineering, The University of Texas at Austin, USA}
\affiliation{Wyant College of Optical Sciences, The University of Arizona, USA}
\author{Chaohan Cui}
\affiliation{Wyant College of Optical Sciences, The University of Arizona, USA}
\author{Yongzhou Xue}
\affiliation{Wyant College of Optical Sciences, The University of Arizona, USA}
\author{Paokang Chen}
\affiliation{Chandra Department of Electrical and Computer Engineering, The University of Texas at Austin, USA}
\affiliation{Wyant College of Optical Sciences, The University of Arizona, USA}
\author{Linran Fan}
\email{linran.fan@utexas.edu}
\affiliation{Chandra Department of Electrical and Computer Engineering, The University of Texas at Austin, USA}

\begin{abstract}


The interaction between photons and phonons plays a crucial role in broad areas ranging from optical sources and modulators to quantum transduction and metrology. 
The performance can be further improved using integrated photonic-phononic devices, promising enhanced interaction strength and large-scale integration.
While the enhanced interaction has been widely demonstrated, it is challenging to realize large-scale integrated photonic-phononic circuits due to material limitations. 
Here, we resolve this critical issue by using gallium nitride on sapphire for scalable photonic-phononic integrated circuits. Both optical and acoustic fields are confined in sub-wavelength scales without suspended structures. This enables us to achieve the efficient launching, flexible routing, and reconfigruable processing of optical and acoustic fields simultaneously. 
With the controlled photonic-phononic interaction and strong piezoelectric effect, we further demonstrate the reconfigurable conversion between frequency-multiplexed RF and optical signals mediated by acoustics. This work provides an ideal platform for achieving ultimate performance of photonic-phononic hybrid systems with high efficiency, multiple functions, and large scalability. 

\end{abstract}

 \maketitle

\textbf{Introduction}


Engineering hybrid systems, combining complementary features from different domains, have the potential to achieve performance significantly surpassing individual systems. One iconic example is hybrid electronic-photonic systems, which are widely used in communication systems and show great promise to improve the speed and efficiency of machine learning tasks~\cite{feng2022compact,zhang2020scalable,shastri2021photonics}. 
Besides electronics and photonics, acoustics (phononics) also plays a critical role in modern information systems, widely used for RF signal processing~\cite{yamagata2022surface}, navigation~\cite{munafo2017acoustic}, and sensing~\cite{liu2016surface}. Recently, significant efforts have been devoted to the development of hybrid photonic-phononic systems, combining the quantum-limited detection and large bandwidth benefits of optics with high sensitivity and strong RF interaction advantages of acoustics~\cite{safavi2019controlling,shin2015control}. This has led to novel functions and improved performance in critical applications such as quantum transduction~\cite{mirhosseini2020superconducting,chen2023optomechanical}, inertia sensing~\cite{huang2020chip}, and optical modulation~\cite{tadesse2014sub,zhang2024integrated}. 
However, the development of hybrid photonic-phononic systems is still in infancy compared with hybrid electronic-photonic systems. 
Large-scale electronic-photonic integrated circuits have been widely demonstrated for advanced information processing tasks~\cite{feng2022compact,zhang2020scalable,shastri2021photonics,ying2020electronic,mukherjee2021towards,behroozpour2016electronic}.
In contrast, hybrid photonic-phononic systems are mostly restricted to single devices with basic functions.

\begin{figure*}[htbp]
\centering
\includegraphics[width=\textwidth]{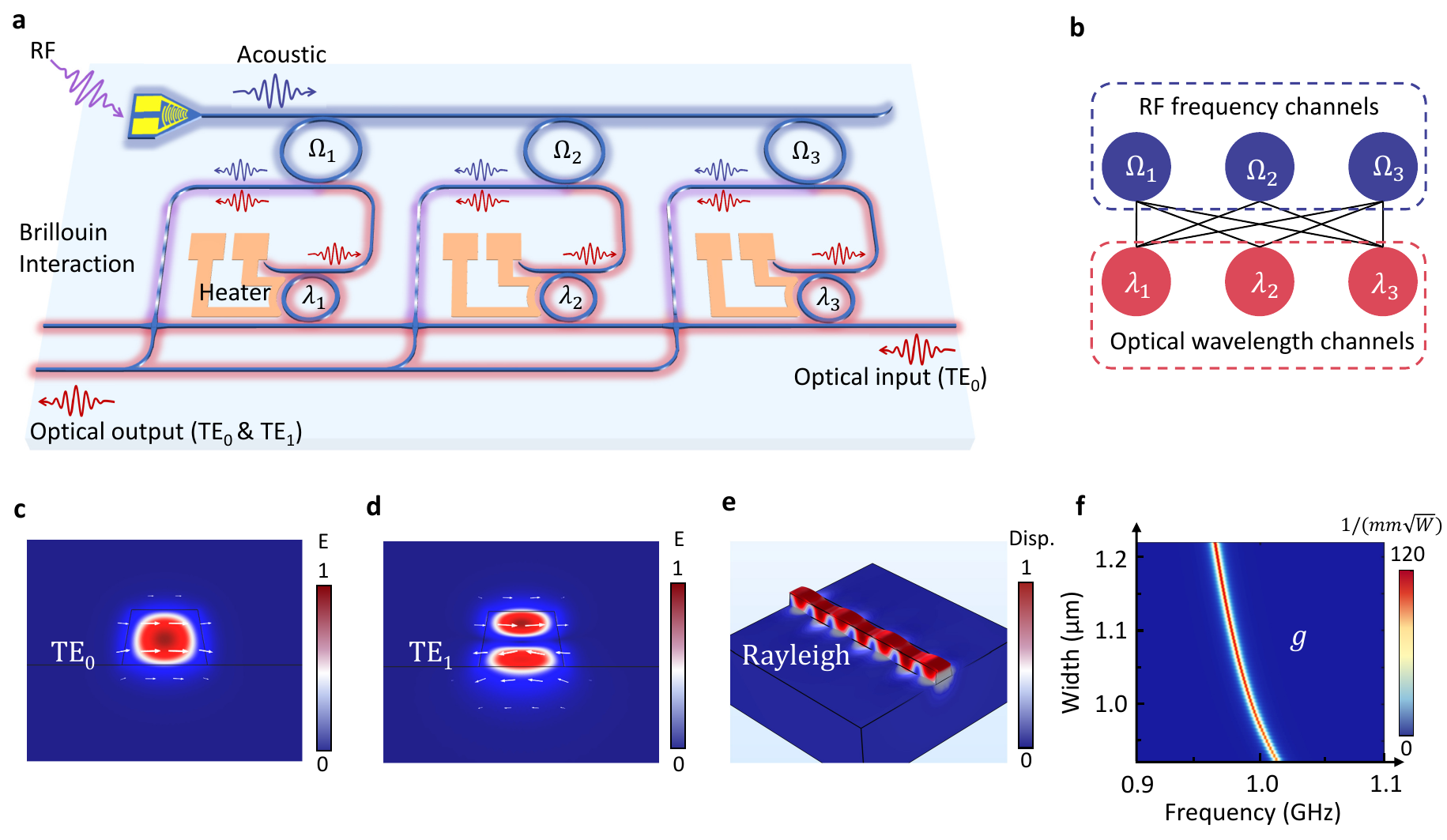}
\caption{\textbf{Photonic-phononic integrated circuits}. \textbf{a}, Schematic of the hybrid photonic-photonic integrated circuit for multiplexed signal conversion between RF and optical fields mediated by acoustic fields. RF, optical, and acoustic fields are labeled in purple, red, and blue respectively. Different optical wavelength channels ($\lambda_1$, $\lambda_2$, $\lambda_3$) are selected by tunable optical ring resonators. Different RF frequency channels ($\Omega_1$, $\Omega_2$, $\Omega_3$) are selected by acoustic ring resonators.
\textbf{b}, Reconfiguration diagram between different RF frequency and optical wavelength channels.
\textbf{c}, field profile of the optical input TE$_0$ mode with electric field direction arrows.
\textbf{d}, field profile of the optical output TE$_1$ mode with electric field direction arrows.
\textbf{e}, the displacement profile of the acoustic fundamental Rayleigh-like mode.
\textbf{f}, photonic-phononic coupling strength as a function of the waveguide width and acoustic frequency.}
\label{fig:figure1}
\end{figure*}

The simultaneous confinement of optical and acoustic fields is required to realize large-scale photonic-phononic integrated circuits. The confinement of optical fields can be easily achieved because waveguide materials typically have lower speed of light than cladding materials.
Similarly, the confinement of acoustic fields requires that waveguide materials also have smaller acoustic velocities~\cite{laude2021principles,wang2024perspectives}, which most integrated platforms do not satisfy~\cite{fu2019phononic}. Consequently, suspended structures are typically used to achieve the simultaneous confinement of optical and acoustic fields~\cite{shao2019microwave,shin2013tailorable}. This inevitably causes critical issues in device robustness and layout flexibility. Thus, the demonstration of large-scale photonic-phononic integrated circuit remains elusive. Besides the simultaneous confinement of optical and acoustic fields, it is also beneficial to use piezoelectric materials, which can couple strongly with RF fields for the efficient excitation of acoustic fields. Moreover, it is highly preferred that the material platform is compatible with standard wafer-scale semiconductor fabrication processing.

    
    
In this work, we develop a scalable platform for photonic-phonoinc integrated circuits based on gallium nitride (GaN) on sapphire. GaN has a significantly higher optical refractive index than sapphire. Notably, the velocities of both transverse and longitudinal acoustic fields in GaN are remarkably lower than sapphire ~\cite{fu2019phononic}. Therefore, the simultaneous confinement of both optical and acoustic fields can be realized in GaN waveguides on sapphire substrates without using suspended structures. 
To demonstrate the scalability of this platform, we fabricate a large-scale photonic-phononic integrated circuit consisted of diverse functional blocks including signal launching, routing, combining, de/multiplexing, in both optical and acoustic domains. As both optical and acoustic fields are confined in sub-wavelength scales, the photonic-phononic interaction strength is also enhanced. Further leveraging the piezoelectric effect of GaN, we realize the reconfigurable signal conversion between different RF frequency channels and optical wavelength channels mediated by acoustic fields.

\maketitle
\vspace{4pt}

\textbf{Results}

The photonic-phononic integrated circuit is depicted schematically in Fig.\;\ref{fig:figure1}a. RF signals with multiple carrier frequencies (violet) are converted into acoustic fields (blue) using integrated interdigital transducers (IDTs) via the piezoelectric effect~\cite{fu2019phononic,xu2022high}. Subsequently,  acoustic fields are focused into the waveguide with a sub-wavelength cross-section. An array of acoustic ring resonators are evanescently coupled to the waveguide to separate different RF frequency channels in the acoustic domain. 
Optical fields (red) are launched into the waveguide through edge coupling. An array of optical ring resonators are used to separate different wavelengths. We use the add-drop configuration for both acoustic and optical ring resonators, and the output waveguides are shared by acoustic and optical fields for photonic-phononic interactions. 
Thermo-optic heaters are integrated to control resonant wavelengths of optical ring resonators. By applying different electric currents in thermo-optic heaters, we can reconfigure the interaction between optical and acoustic fields, thus realize the arbitrary mapping from RF frequency channels to optical wavelength channels (Fig.\;\ref{fig:figure1}b).

The photonic-phononic interaction is implemented based on the Brillouin scattering process, with the coupling coefficient ($g$) determined by optical and acoustic mode profiles~\cite{wiederhecker2019brillouin,wolff2015stimulated}. 
We choose the fundamental Rayleigh-like mode for the acoustic field~\cite{hashimoto2000surface}, because the dominant displacement is out-of-plane, thus can be excited most efficiently using the largest $d_{33}$ piezoelectric coefficient (Fig.\;\ref{fig:figure1}e) ~\cite{xu2022high}. 
Considering the anti-symmetric out-of-plane displacement of the fundamental Rayleigh-like acoustic mode, we choose the fundamental and first-order transverse-electric (TE$_0$ and TE$_1$) modes as the optical input and output respectively~(Fig.\;\ref{fig:figure1}c and d)~\cite{zhang2024integrated}.
Due to the strong sub-wavelength confinement of both optical and acoustic fields, the Brillouin scattering process shows significant geometric dispersion~(Fig.\;\ref{fig:figure1}f). 
For GaN waveguides with height 0.95~$\mu$m and width 1~$\mu$m, the largest acousto-optic coupling coefficient $g\approx120\;(\mathrm{mm} \mathrm{W}^{1/2})^{-1}$ is achieved, enabling the strong photonic-phononic interaction between the acoustic frequency $\Omega/2\pi\approx0.98$\;GHz and optical input wavelength 1550 nm. Our Brillouin coupling strength $g$ is on par with the reported best result from suspended lithium niobate (LiNbO$_3$) platform (377 ($\mathrm{mm W^{1/2}}$)$^{-1}$)~\cite{sarabalis2021acousto}.

\begin{figure}[htbp]
\centering
\includegraphics[width = \linewidth]{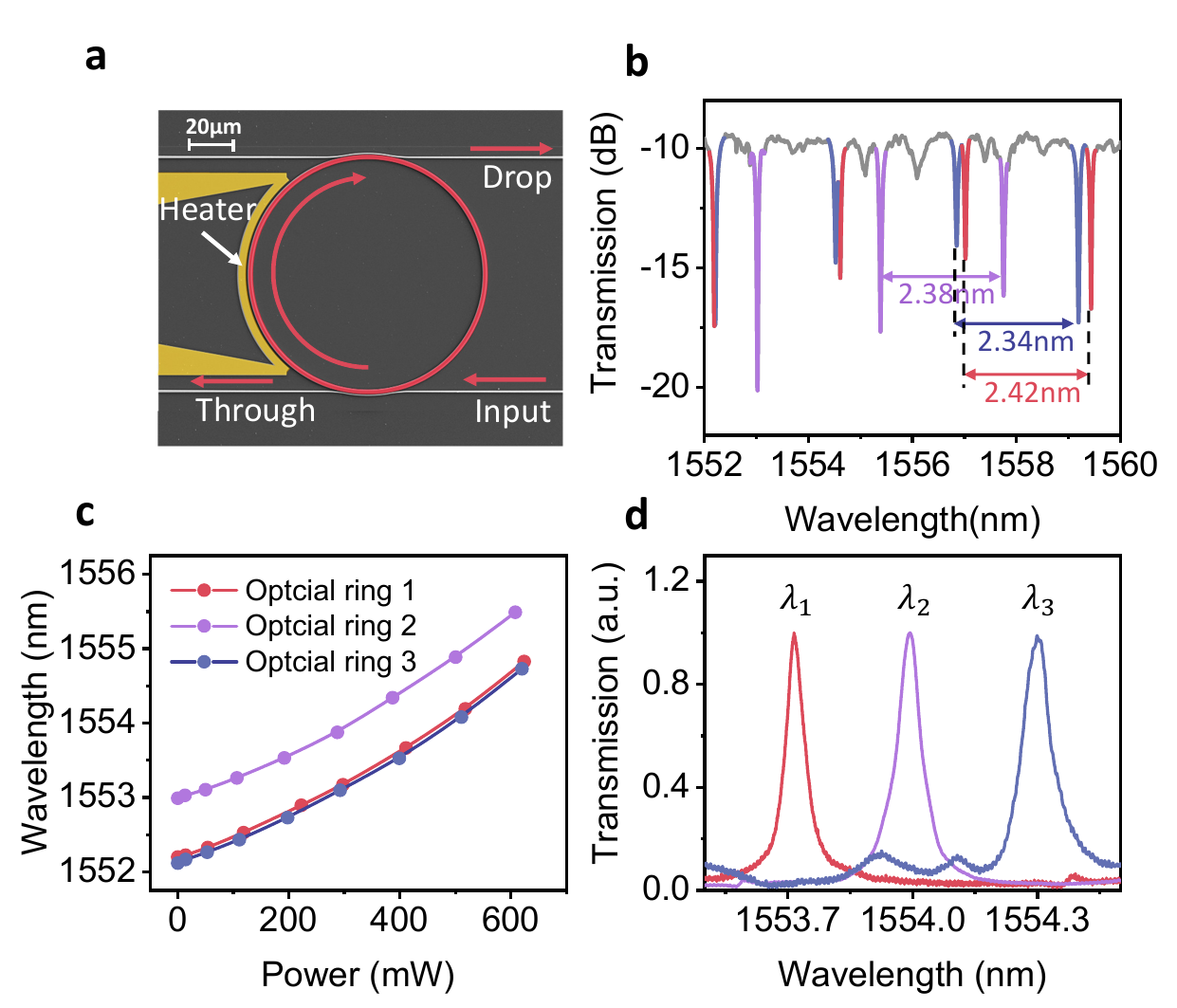}
\caption{ \textbf{Optical performance characterization}. \textbf{a}, False-color scanning electron microscope (SEM) image of an optical ring resonator (red) and a thermo-optic heater (yellow).
\textbf{b}, Through-port transmission with three optical ring resonators with resonances labeled in blue, red, and, purple respectively. Corresponding free-spectral ranges are labeled for each optical ring resonator.
\textbf{c}, Resonant wavelength shifts of the three optical ring resonators under thermo-optic tuning.
\textbf{d}, Drop-port transmission of the three optical ring resonators with resonances tuned at 1553.7 nm ($\lambda_1$), 1554.0 nm ($\lambda_2$), and 1554.3 nm ($\lambda_3$) respectively.
}
\label{fig:figure2}
\end{figure}

\begin{figure}[tb]
\centering
\includegraphics[width=\linewidth]{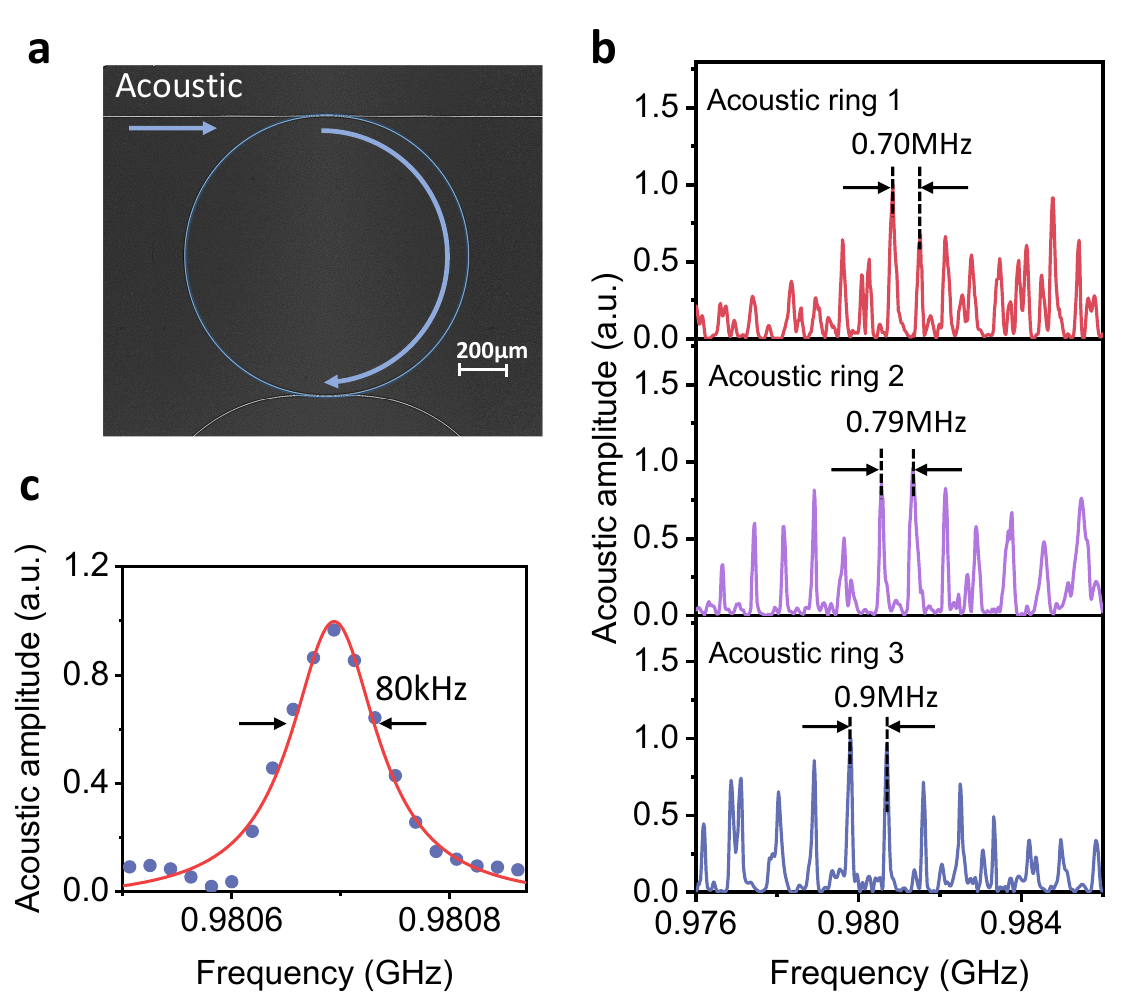}
\caption{\textbf{Acoustic performance characterization}. \textbf{a}, False-color scanning electron microscope (SEM) image of an acoustic ring resonator \textbf{b}, Transmission spectrum of acoustic ring resonators measured with optical detection. Corresponding free-spectral ranges are labeled for each acoustic ring resonator. \textbf{c}, One resonance of the third acoustic ring resonator with Lorentz fitting, showing a quality factor over $12\times10^3$.
}
\label{fig:figure3}
\end{figure}

\begin{figure*}[htbp]
\centering
\includegraphics[width=\textwidth]{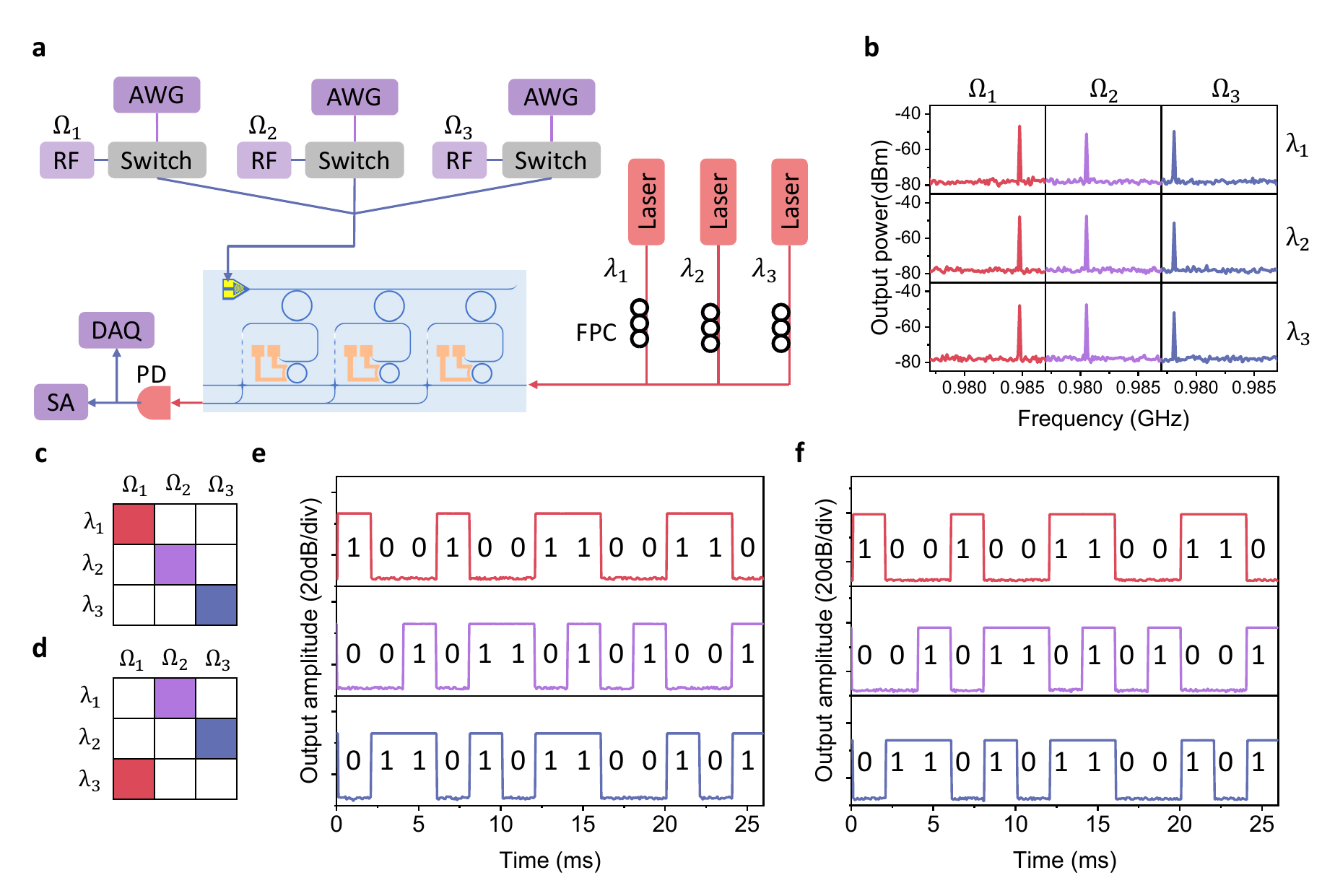}
\caption{\textbf{Reconfigurable signal conversion between RF and optical fields mediated by acoustic fields.}
\textbf{a}, Schematic of the measurement setup. AWG, arbitrary wave generator; SA, spectrum analyzer; PD, photodiode; DAQ, data acquisition card. \textbf{b}, Arbitrarily reconfigurable mapping between RF frequencies ($\Omega_1$, $\Omega_2$, $\Omega_3$) and optical wavelengths ($\lambda_1$, $\lambda_2$, $\lambda_3$) with extinction ratio over 30 dB.
\textbf{c} and \textbf{d}, two specific mapping diagrams from RF frequencies to optical wavelengths.
\textbf{e} and \textbf{f}, Measured data stream after signal conversion from RF frequencies to optical wavelengths with the mapping diagram in \textbf{c} and \textbf{d} respectively. Input bits (0 and 1) are labeled for reference.
}
\label{fig:figure4}
\end{figure*}

The device is fabricated with c-plane 0.95-$\mu$m thick GaN template on sapphire grown by MOCVD (Supplementary Section 1). We first characterize the optical performance of the photonic-phononic integrated circuit (Fig.\;\ref{fig:figure2}a). Three optical ring resonators with nominal width of 2 $\mu$m and radii of 65 $\mu$m, 66 $\mu$m and 67 $\mu$m are evanescently coupled to the same waveguide.
Three groups of resonances are clearly observed in the transmission spectrum (marked by different colors in Fig.\;\ref{fig:figure2}b). The corresponding free spectral ranges are 2.34 nm, 2.38 nm, and 2.42 nm respectively, matching the simulation result of TE$_0$ optical modes (Supplementary Section 2) . 
The average linewidth of optical resonances is 3~GHz, which is sufficient for dense wavelength-division multiplexing in the optical domain. Assuming under-coupled conditions, the coupling between optical resonances and the bus waveguide can be estimated as 0.53~GHz.
Optical ring resonators can be tuned by thermo-optic heaters with minimized crosstalks (Supplementary Section 2). The tuning efficiency is measured around 4~pm/mW (Fig.\;\ref{fig:figure2}c). 
In this work, three wavelength channels at 1553.7~nm ($\lambda_1$), 1554.0~nm ($\lambda_2$), and 1554.3~nm ($\lambda_3$) are selected~(Fig.\;\ref{fig:figure2}d).

Next, we characterize the performance of acoustic ring resonators (Fig.~\ref{fig:figure3}a). Acoustic fields are launched through focused IDTs (Supplementary Section 3), and detected with optical fields after photonic-phononic interaction~\cite{zhang2024integrated}.
The acoustic ring resonators have a waveguide width of 4~$\mu$m and radii of 700~$\mu$m, 800~$\mu$m, and 900~$\mu$m respectively.
The free spectral ranges are measured as 0.7 MHz, 0.79 MHz, and 0.9 MHz (Fig. \ref{fig:figure3}b), corresponding to acoustic group velocities of 3958 m/s, 3970 m/s, and 3958 m/s, respectively.
These results match with the simulated value of 3981 m/s for the fundamental Rayleigh-like mode. The slight deviation between simulated and measured values could be attributed to the use of 2D simulation (supplementary section 4) or the uncertainty during device fabrication. This result confirms that the acoustic fields maintain in the fundamental Rayleigh-like mode after coupling from the waveguide into the ring resonators. Noteworthy, there exists other peaks, indicating the excitation of other acoustic modes. They could be excited directly by IDT or transferred from fundamental Rayleigh-like fields due to mode hybridization in the taper regime between IDT and acoustic waveguides. The fundamental Rayleigh-like fields in waveguides could also excite high-order modes in acoustic rings.
The quality factors of the acoustic ring resonators range from $5\times10^3$ to $12\times10^3$~(Fig.~\ref{fig:figure3}c), corresponding to acoustic propagation loss between 1.3 dB/mm and 0.56 dB/mm (Supplementary section 4). The average coupling strength between the acoustic ring and input waveguide is estimated around 1.47 MHz with numerical simulation (Supplementary section 5).

After confirming the performance of optical and acoustic circuits separately, we implement the reconfigurable signal conversion between different RF frequency channels and optical wavelength channels mediated by acoustic fields (Fig.\;\ref{fig:figure4}a). RF signals at frequencies 0.985~GHz ($\Omega_1$), 0.980 GHz ($\Omega_2$), and 0.978 GHz ($\Omega_3$) are generated and launched into the acoustic waveguide through the IDT. The RF power is fixed at 20~dBm. These frequencies are chosen to match the resonant frequencies of the three acoustic ring resonators.
At the same time, three optical pumps at 1553.7~nm ($\lambda_1$), 1554.0~nm ($\lambda_2$), and 1554.3~nm ($\lambda_3$) are generated by tunable lasers, and coupled to the optical input port. The optical pump power is 5~mW.
By tuning the on-chip thermo-optic heaters, we can control which optical ring resonators optical pumps can couple into. Therefore, optical pumps can be reconfigured to interact with acoustic fields at different frequencies (Fig.\;\ref{fig:figure1}b). Arbitrary combination between acoustic fields and optical pumps can be realized. Output optical signals are detected by a 1~GHz photodetector, followed by a electronic spectrum analyzer and data acquisition card for frequency- and time-domain measurement respectively. From the frequency-domain measurement, we can clearly see that the crosstalk among different frequency channels is minimum (Fig.\;\ref{fig:figure4}b). Signal-to-noise ratio (SNR) above 30~dB can be achieved. This value is limited by the bandwidth of the spectrum analyzer which sets the noise floor, instead of the intrinsic performance of the photonic-phononic integrated circuit (Supplementary section 6). SNR over 55~dB has been reported in Brillouin active systems with piezoelectric excitation~\cite{otterstrom2019resonantly,shao2019microwave}. We further test the signal conversion in the time domain. 
We use non-return-to-zero on-off keying to encode binary sequence into RF signals using arbitrary waveform generators and switches~\cite{powell2022integrated}. Each RF frequency channel has completely independent data sets. Two different mapping configurations between RF frequency and optical wavelength channels, ($\Omega_1\;\Omega_2\;\Omega_3) \rightarrow (\lambda_1\;\lambda_2\;\lambda_3$) and ($\Omega_1\;\Omega_2\;\Omega_3) \rightarrow (\lambda_3\;\lambda_2\;\lambda_1$), are tested (Fig.\;\ref{fig:figure4}c and d). For both configurations, the high-fidelity signal readout from the optical field is realized with signal-to-noise ratio above 35 dB  (Fig.\;\ref{fig:figure4}e and f).

\begin{figure}[tb]
\centering
\includegraphics[width=\linewidth]{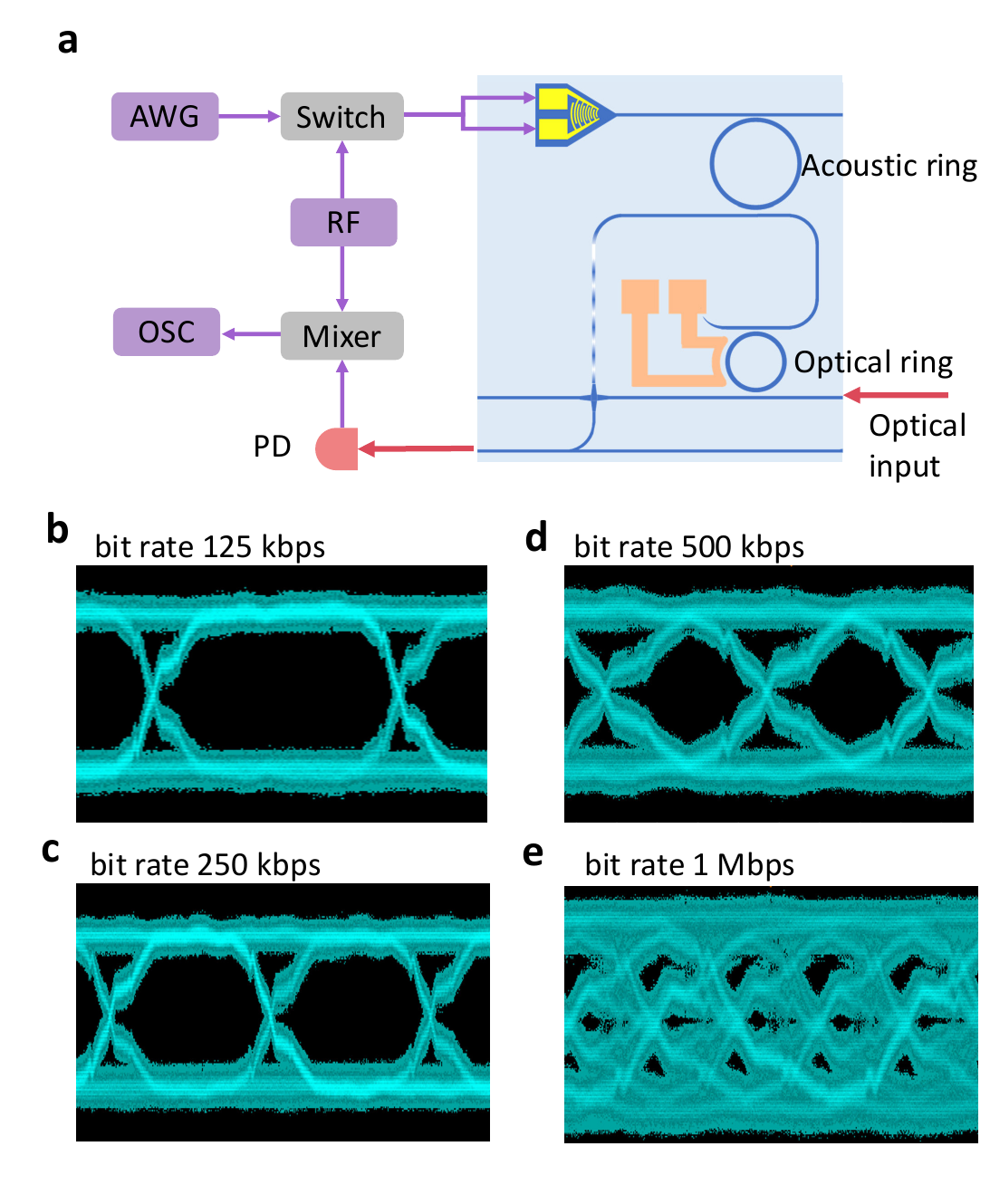}
\caption{\textbf{High-speed performance for signal conversion} 
\textbf{a}, Schematic to measure the eye diagram for the signal conversion from RF to optical fields. AWG, arbitrary wave generator; OSC, oscilloscope. 
\textbf{b-e}, Measured eye diagram with input bit rates at 125~kbps, 250~kbps, 500~kbps, and 1~Mbps respectively}
\label{fig:figure5}
\end{figure}

We further evaluate the high-speed signal conversion performance by measuring the eye diagram at different data rates. The measurement setup is presented in Fig.\;\ref{fig:figure5}a. 
Instead of using DAQ to directly record data, output signals are first demodulated with the carrier frequency to the baseband and recorded with an oscilloscope. Different segments of the data stream are overlay on the oscilloscope by setting the trigger voltage to the middle of the on and off states.
We use the second acoustic ring resonator with radius 800~$\mu$m. We focus on the acoustic resonance at 0.98~GHz with a quality factor of $6\times10^3$.
The optical pump wavelength is 1554~nm, and the input optical pump power is fixed at 10 mW. The resonant frequency of the second optical ring resonator is tuned to 1554~nm to match the input optical wavelength.
Psuedo-random data in the non-return-zero on-off-keying is generated by the arbitrary wave generator.
With a low data rate (125~kbps), the measured bit error rate is below $1\times10^{-6}$ (Fig.\;\ref{fig:figure5}b). As we increase the data rate, the opening of the eye diagram decreases, leading to higher bit error rates (Fig.\;\ref{fig:figure5}c and d). With a data rate of 1~Mbps, the eye diagram is completely closed. In our hybrid photonic-phononinc integrated systems, acoustic resonances have much narrower linewidths than optical resonances and the phase matching window of the Brillouin interaction. Therefore, the maximum data rate is limited by the phonon lifetime of the acoustic ring resonance. From the acoustic resonance quality factor, we can estimate the phonon lifetime around 6~$\mu$s, which matches the achievable data rate.

\vspace{4pt}
\textbf{Discussion}
\vspace{4pt}

There is still significant space to improve the efficiency  of the proposed hybrid photoninc-phononic integrated circuits. 
The bandwidth and efficiency of IDT can be significantly improved by adopting advanced IDT designs, such as unidirectional and chirped IDT structures~\cite{yamanouchi2005ultra,fall2017generation}. Additionally, employing integrated material platforms with stronger piezoelectricity can also lead to improvements in IDT efficiency and bandwidth. For example, LiNbO$_3$ on sapphire has an effective electroacoustic coupling factor of 0.15~\cite{mayor2021gigahertz}, which can lead to over 7-times and 50-times improvements for IDT bandwidth and efficiency, respectively~\cite{hashimoto2000surface}.
The tuning of optical ring resonators can also be implemented with electro-optical effect instead of thermal-optical effect, leading to lower power consumption, higher modulation speed, and smaller crosstalk. While it is sufficient to reconfigure only optical ring resonators in this work, it is noteworthy that acoustic ring resonators can also be tuned using thermal-elastic and electro-elastic effects~\cite{shao2022thermal,shao2022electrical}.
The propagation loss of the photonic circuit can be further improved by using GaN wafers with lower free-carrier density~\cite{chen2017low,zhang2025highly}. These improvements can considerably enhance the device performance and extend the scalability of our system.
Additionally, improvements in power efficiency can enable the operation of the hybrid photonic-phononic integrated circuits at cryogenic temperatures, which can further reduce the acoustic loss. The cryogenic operation will also allow the integration between hybrid photonic-phononic circuits and superconducting devices, leading to potential applications in quantum transduction~\cite{mirhosseini2020superconducting} and quantum phononics ~\cite{ravichandran2020phonon, mayor2021gigahertz,chu2017quantum,manenti2017circuit}.

In conclusion, we developed a scalable platform for hybrid photonic-phononic integrated circuits with gallium nitride on sapphire. Simultaneous confinement of acoustic and optical fields in the sub-wavelength regime can be realized without using suspended structures. This enables the flexible layout of complex hybrid circuits with advanced functions. We fabricated a large-scale hybrid photonic-phononic hybrid circuits to implement the signal conversion between RF and optical fields. Different function blocks including signal launching, routing, combining, de/multiplexing, in both optical and acoustic domains are integrated on the same chip. The capability to reconfigure the signal conversion is also demonstrated.
Furthermore, the GaN-on-sapphire platform has been widely used in power electronics~\cite{zheng2021gallium}, RF electronics~\cite{zhang2020heterogeneously}, and optoelectronic devices ~\cite{baek2023ultra,vafadar2023ultralow,jain2023atomically}. The well-established wafer-scale growth and fabrication process also provides important benefits in the development of large-scale hybrid photonic-phononic integrated circuits.

\vspace{4pt}
\textbf{Methods}

\textbf{Device fabrication}. The device fabrication process is detailed in Supplementary Section 1. Briefly, we pattern the 0.95-$\mu$m-GaN wafer using electron-beam lithography and plasma dry etching. IDTs are fabricated through electron-beam lithography with PMMA resist, followed by the deposition of 5-nm titanium and 100-nm aluminum layers, and finalized by acetone lift-off. The heaters, consisting of 5-nm titanium and 100-nm gold layers, are fabricated using a similar method with IDT.

\textbf{Measurement methods}. Optical ring filters are characterized through transmission measurement using a tunable laser operating in the telecommunication C-band. For acoustic ring filters, we implement a hybrid measurement scheme. RF signals from VNA is sent to IDT to drive acoustic fields. At the same time, a continuous-wave probe laser is launched into the acousto-optic waveguides and modulated by acoustic fields. Then modulated optical fields are detected by the photodetector whose output is sent back to VNA.

\vspace{4pt}
\textbf{Data availability}

The Source Data underlying the figures of this study are available with the paper. All raw data generated during the current study are available from the corresponding authors upon request.

\bibliography{main_text}

\begin{thebibliography}{43}%
\makeatletter
\providecommand \@ifxundefined [1]{%
 \@ifx{#1\undefined}
}%
\providecommand \@ifnum [1]{%
 \ifnum #1\expandafter \@firstoftwo
 \else \expandafter \@secondoftwo
 \fi
}%
\providecommand \@ifx [1]{%
 \ifx #1\expandafter \@firstoftwo
 \else \expandafter \@secondoftwo
 \fi
}%
\providecommand \natexlab [1]{#1}%
\providecommand \enquote  [1]{``#1''}%
\providecommand \bibnamefont  [1]{#1}%
\providecommand \bibfnamefont [1]{#1}%
\providecommand \citenamefont [1]{#1}%
\providecommand \href@noop [0]{\@secondoftwo}%
\providecommand \href [0]{\begingroup \@sanitize@url \@href}%
\providecommand \@href[1]{\@@startlink{#1}\@@href}%
\providecommand \@@href[1]{\endgroup#1\@@endlink}%
\providecommand \@sanitize@url [0]{\catcode `\\12\catcode `\$12\catcode `\&12\catcode `\#12\catcode `\^12\catcode `\_12\catcode `\%12\relax}%
\providecommand \@@startlink[1]{}%
\providecommand \@@endlink[0]{}%
\providecommand \url  [0]{\begingroup\@sanitize@url \@url }%
\providecommand \@url [1]{\endgroup\@href {#1}{\urlprefix }}%
\providecommand \urlprefix  [0]{URL }%
\providecommand \Eprint [0]{\href }%
\providecommand \doibase [0]{http://dx.doi.org/}%
\providecommand \selectlanguage [0]{\@gobble}%
\providecommand \bibinfo  [0]{\@secondoftwo}%
\providecommand \bibfield  [0]{\@secondoftwo}%
\providecommand \translation [1]{[#1]}%
\providecommand \BibitemOpen [0]{}%
\providecommand \bibitemStop [0]{}%
\providecommand \bibitemNoStop [0]{.\EOS\space}%
\providecommand \EOS [0]{\spacefactor3000\relax}%
\providecommand \BibitemShut  [1]{\csname bibitem#1\endcsname}%
\let\auto@bib@innerbib\@empty
\bibitem [{\citenamefont {Feng}\ \emph {et~al.}(2022)\citenamefont {Feng}, \citenamefont {Gu}, \citenamefont {Zhu}, \citenamefont {Ying}, \citenamefont {Zhao}, \citenamefont {Pan},\ and\ \citenamefont {Chen}}]{feng2022compact}%
  \BibitemOpen
  \bibfield  {author} {\bibinfo {author} {\bibfnamefont {Chenghao}\ \bibnamefont {Feng}}, \bibinfo {author} {\bibfnamefont {Jiaqi}\ \bibnamefont {Gu}}, \bibinfo {author} {\bibfnamefont {Hanqing}\ \bibnamefont {Zhu}}, \bibinfo {author} {\bibfnamefont {Zhoufeng}\ \bibnamefont {Ying}}, \bibinfo {author} {\bibfnamefont {Zheng}\ \bibnamefont {Zhao}}, \bibinfo {author} {\bibfnamefont {David~Z}\ \bibnamefont {Pan}}, \ and\ \bibinfo {author} {\bibfnamefont {Ray~T}\ \bibnamefont {Chen}},\ }\bibfield  {title} {\enquote {\bibinfo {title} {A compact butterfly-style silicon photonic--electronic neural chip for hardware-efficient deep learning},}\ }\href@noop {} {\bibfield  {journal} {\bibinfo  {journal} {Acs Photonics}\ }\textbf {\bibinfo {volume} {9}},\ \bibinfo {pages} {3906--3916} (\bibinfo {year} {2022})}\BibitemShut {NoStop}%
\bibitem [{\citenamefont {Zhang}\ \emph {et~al.}(2020{\natexlab{a}})\citenamefont {Zhang}, \citenamefont {Samanta}, \citenamefont {Shang},\ and\ \citenamefont {Yoo}}]{zhang2020scalable}%
  \BibitemOpen
  \bibfield  {author} {\bibinfo {author} {\bibfnamefont {Yu}~\bibnamefont {Zhang}}, \bibinfo {author} {\bibfnamefont {Anirban}\ \bibnamefont {Samanta}}, \bibinfo {author} {\bibfnamefont {Kuanping}\ \bibnamefont {Shang}}, \ and\ \bibinfo {author} {\bibfnamefont {SJ~Ben}\ \bibnamefont {Yoo}},\ }\bibfield  {title} {\enquote {\bibinfo {title} {Scalable 3d silicon photonic electronic integrated circuits and their applications},}\ }\href@noop {} {\bibfield  {journal} {\bibinfo  {journal} {IEEE Journal of Selected Topics in Quantum Electronics}\ }\textbf {\bibinfo {volume} {26}},\ \bibinfo {pages} {1--10} (\bibinfo {year} {2020}{\natexlab{a}})}\BibitemShut {NoStop}%
\bibitem [{\citenamefont {Shastri}\ \emph {et~al.}(2021)\citenamefont {Shastri}, \citenamefont {Tait}, \citenamefont {Ferreira~de Lima}, \citenamefont {Pernice}, \citenamefont {Bhaskaran}, \citenamefont {Wright},\ and\ \citenamefont {Prucnal}}]{shastri2021photonics}%
  \BibitemOpen
  \bibfield  {author} {\bibinfo {author} {\bibfnamefont {Bhavin~J}\ \bibnamefont {Shastri}}, \bibinfo {author} {\bibfnamefont {Alexander~N}\ \bibnamefont {Tait}}, \bibinfo {author} {\bibfnamefont {Thomas}\ \bibnamefont {Ferreira~de Lima}}, \bibinfo {author} {\bibfnamefont {Wolfram~HP}\ \bibnamefont {Pernice}}, \bibinfo {author} {\bibfnamefont {Harish}\ \bibnamefont {Bhaskaran}}, \bibinfo {author} {\bibfnamefont {C~David}\ \bibnamefont {Wright}}, \ and\ \bibinfo {author} {\bibfnamefont {Paul~R}\ \bibnamefont {Prucnal}},\ }\bibfield  {title} {\enquote {\bibinfo {title} {Photonics for artificial intelligence and neuromorphic computing},}\ }\href@noop {} {\bibfield  {journal} {\bibinfo  {journal} {Nature Photonics}\ }\textbf {\bibinfo {volume} {15}},\ \bibinfo {pages} {102--114} (\bibinfo {year} {2021})}\BibitemShut {NoStop}%
\bibitem [{\citenamefont {Yamagata}\ \emph {et~al.}(2022)\citenamefont {Yamagata}, \citenamefont {Cao}, \citenamefont {John},\ and\ \citenamefont {Hashemi}}]{yamagata2022surface}%
  \BibitemOpen
  \bibfield  {author} {\bibinfo {author} {\bibfnamefont {Masashi}\ \bibnamefont {Yamagata}}, \bibinfo {author} {\bibfnamefont {Ning}\ \bibnamefont {Cao}}, \bibinfo {author} {\bibfnamefont {Demis~D}\ \bibnamefont {John}}, \ and\ \bibinfo {author} {\bibfnamefont {Hossein}\ \bibnamefont {Hashemi}},\ }\bibfield  {title} {\enquote {\bibinfo {title} {Surface-acoustic-wave waveguides for radio frequency signal processing},}\ }\href@noop {} {\bibfield  {journal} {\bibinfo  {journal} {IEEE Transactions on Microwave Theory and Techniques}\ }\textbf {\bibinfo {volume} {71}},\ \bibinfo {pages} {931--944} (\bibinfo {year} {2022})}\BibitemShut {NoStop}%
\bibitem [{\citenamefont {Munafo}\ and\ \citenamefont {Ferri}(2017)}]{munafo2017acoustic}%
  \BibitemOpen
  \bibfield  {author} {\bibinfo {author} {\bibfnamefont {Andrea}\ \bibnamefont {Munafo}}\ and\ \bibinfo {author} {\bibfnamefont {Gabriele}\ \bibnamefont {Ferri}},\ }\bibfield  {title} {\enquote {\bibinfo {title} {An acoustic network navigation system},}\ }\href@noop {} {\bibfield  {journal} {\bibinfo  {journal} {Journal of Field Robotics}\ }\textbf {\bibinfo {volume} {34}},\ \bibinfo {pages} {1332--1351} (\bibinfo {year} {2017})}\BibitemShut {NoStop}%
\bibitem [{\citenamefont {Liu}\ \emph {et~al.}(2016)\citenamefont {Liu}, \citenamefont {Chen}, \citenamefont {Cai}, \citenamefont {Ali}, \citenamefont {Tian}, \citenamefont {Tao}, \citenamefont {Yang},\ and\ \citenamefont {Ren}}]{liu2016surface}%
  \BibitemOpen
  \bibfield  {author} {\bibinfo {author} {\bibfnamefont {Bo}~\bibnamefont {Liu}}, \bibinfo {author} {\bibfnamefont {Xiao}\ \bibnamefont {Chen}}, \bibinfo {author} {\bibfnamefont {Hualin}\ \bibnamefont {Cai}}, \bibinfo {author} {\bibfnamefont {Mohammad~Mohammad}\ \bibnamefont {Ali}}, \bibinfo {author} {\bibfnamefont {Xiangguang}\ \bibnamefont {Tian}}, \bibinfo {author} {\bibfnamefont {Luqi}\ \bibnamefont {Tao}}, \bibinfo {author} {\bibfnamefont {Yi}~\bibnamefont {Yang}}, \ and\ \bibinfo {author} {\bibfnamefont {Tianling}\ \bibnamefont {Ren}},\ }\bibfield  {title} {\enquote {\bibinfo {title} {Surface acoustic wave devices for sensor applications},}\ }\href@noop {} {\bibfield  {journal} {\bibinfo  {journal} {Journal of semiconductors}\ }\textbf {\bibinfo {volume} {37}},\ \bibinfo {pages} {021001} (\bibinfo {year} {2016})}\BibitemShut {NoStop}%
\bibitem [{\citenamefont {Safavi-Naeini}\ \emph {et~al.}(2019)\citenamefont {Safavi-Naeini}, \citenamefont {Van~Thourhout}, \citenamefont {Baets},\ and\ \citenamefont {Van~Laer}}]{safavi2019controlling}%
  \BibitemOpen
  \bibfield  {author} {\bibinfo {author} {\bibfnamefont {Amir~H}\ \bibnamefont {Safavi-Naeini}}, \bibinfo {author} {\bibfnamefont {Dries}\ \bibnamefont {Van~Thourhout}}, \bibinfo {author} {\bibfnamefont {Roel}\ \bibnamefont {Baets}}, \ and\ \bibinfo {author} {\bibfnamefont {Rapha{\"e}l}\ \bibnamefont {Van~Laer}},\ }\bibfield  {title} {\enquote {\bibinfo {title} {Controlling phonons and photons at the wavelength scale: integrated photonics meets integrated phononics},}\ }\href@noop {} {\bibfield  {journal} {\bibinfo  {journal} {Optica}\ }\textbf {\bibinfo {volume} {6}},\ \bibinfo {pages} {213--232} (\bibinfo {year} {2019})}\BibitemShut {NoStop}%
\bibitem [{\citenamefont {Shin}\ \emph {et~al.}(2015)\citenamefont {Shin}, \citenamefont {Cox}, \citenamefont {Jarecki}, \citenamefont {Starbuck}, \citenamefont {Wang},\ and\ \citenamefont {Rakich}}]{shin2015control}%
  \BibitemOpen
  \bibfield  {author} {\bibinfo {author} {\bibfnamefont {Heedeuk}\ \bibnamefont {Shin}}, \bibinfo {author} {\bibfnamefont {Jonathan~A}\ \bibnamefont {Cox}}, \bibinfo {author} {\bibfnamefont {Robert}\ \bibnamefont {Jarecki}}, \bibinfo {author} {\bibfnamefont {Andrew}\ \bibnamefont {Starbuck}}, \bibinfo {author} {\bibfnamefont {Zheng}\ \bibnamefont {Wang}}, \ and\ \bibinfo {author} {\bibfnamefont {Peter~T}\ \bibnamefont {Rakich}},\ }\bibfield  {title} {\enquote {\bibinfo {title} {Control of coherent information via on-chip photonic--phononic emitter--receivers},}\ }\href@noop {} {\bibfield  {journal} {\bibinfo  {journal} {Nature communications}\ }\textbf {\bibinfo {volume} {6}},\ \bibinfo {pages} {6427} (\bibinfo {year} {2015})}\BibitemShut {NoStop}%
\bibitem [{\citenamefont {Mirhosseini}\ \emph {et~al.}(2020)\citenamefont {Mirhosseini}, \citenamefont {Sipahigil}, \citenamefont {Kalaee},\ and\ \citenamefont {Painter}}]{mirhosseini2020superconducting}%
  \BibitemOpen
  \bibfield  {author} {\bibinfo {author} {\bibfnamefont {Mohammad}\ \bibnamefont {Mirhosseini}}, \bibinfo {author} {\bibfnamefont {Alp}\ \bibnamefont {Sipahigil}}, \bibinfo {author} {\bibfnamefont {Mahmoud}\ \bibnamefont {Kalaee}}, \ and\ \bibinfo {author} {\bibfnamefont {Oskar}\ \bibnamefont {Painter}},\ }\bibfield  {title} {\enquote {\bibinfo {title} {Superconducting qubit to optical photon transduction},}\ }\href@noop {} {\bibfield  {journal} {\bibinfo  {journal} {Nature}\ }\textbf {\bibinfo {volume} {588}},\ \bibinfo {pages} {599--603} (\bibinfo {year} {2020})}\BibitemShut {NoStop}%
\bibitem [{\citenamefont {Chen}\ \emph {et~al.}(2023)\citenamefont {Chen}, \citenamefont {Li}, \citenamefont {Lee}, \citenamefont {Chakravarthi}, \citenamefont {Fu},\ and\ \citenamefont {Li}}]{chen2023optomechanical}%
  \BibitemOpen
  \bibfield  {author} {\bibinfo {author} {\bibfnamefont {I-Tung}\ \bibnamefont {Chen}}, \bibinfo {author} {\bibfnamefont {Bingzhao}\ \bibnamefont {Li}}, \bibinfo {author} {\bibfnamefont {Seokhyeong}\ \bibnamefont {Lee}}, \bibinfo {author} {\bibfnamefont {Srivatsa}\ \bibnamefont {Chakravarthi}}, \bibinfo {author} {\bibfnamefont {Kai-Mei}\ \bibnamefont {Fu}}, \ and\ \bibinfo {author} {\bibfnamefont {Mo}~\bibnamefont {Li}},\ }\bibfield  {title} {\enquote {\bibinfo {title} {Optomechanical ring resonator for efficient microwave-optical frequency conversion},}\ }\href@noop {} {\bibfield  {journal} {\bibinfo  {journal} {Nature Communications}\ }\textbf {\bibinfo {volume} {14}},\ \bibinfo {pages} {7594} (\bibinfo {year} {2023})}\BibitemShut {NoStop}%
\bibitem [{\citenamefont {Huang}\ \emph {et~al.}(2020)\citenamefont {Huang}, \citenamefont {Flor~Flores}, \citenamefont {Li}, \citenamefont {Wang}, \citenamefont {Wang}, \citenamefont {Goldberg}, \citenamefont {Zheng}, \citenamefont {Yu}, \citenamefont {Lu}, \citenamefont {Kutzer} \emph {et~al.}}]{huang2020chip}%
  \BibitemOpen
  \bibfield  {author} {\bibinfo {author} {\bibfnamefont {Yongjun}\ \bibnamefont {Huang}}, \bibinfo {author} {\bibfnamefont {Jaime~Gonzalo}\ \bibnamefont {Flor~Flores}}, \bibinfo {author} {\bibfnamefont {Ying}\ \bibnamefont {Li}}, \bibinfo {author} {\bibfnamefont {Wenting}\ \bibnamefont {Wang}}, \bibinfo {author} {\bibfnamefont {Di}~\bibnamefont {Wang}}, \bibinfo {author} {\bibfnamefont {Noam}\ \bibnamefont {Goldberg}}, \bibinfo {author} {\bibfnamefont {Jiangjun}\ \bibnamefont {Zheng}}, \bibinfo {author} {\bibfnamefont {Mingbin}\ \bibnamefont {Yu}}, \bibinfo {author} {\bibfnamefont {Ming}\ \bibnamefont {Lu}}, \bibinfo {author} {\bibfnamefont {Michael}\ \bibnamefont {Kutzer}},  \emph {et~al.},\ }\bibfield  {title} {\enquote {\bibinfo {title} {A chip-scale oscillation-mode optomechanical inertial sensor near the thermodynamical limits},}\ }\href@noop {} {\bibfield  {journal} {\bibinfo  {journal} {Laser \& Photonics Reviews}\ }\textbf {\bibinfo {volume} {14}},\ \bibinfo {pages} {1800329} (\bibinfo {year}
  {2020})}\BibitemShut {NoStop}%
\bibitem [{\citenamefont {Tadesse}\ and\ \citenamefont {Li}(2014)}]{tadesse2014sub}%
  \BibitemOpen
  \bibfield  {author} {\bibinfo {author} {\bibfnamefont {Semere~Ayalew}\ \bibnamefont {Tadesse}}\ and\ \bibinfo {author} {\bibfnamefont {Mo}~\bibnamefont {Li}},\ }\bibfield  {title} {\enquote {\bibinfo {title} {Sub-optical wavelength acoustic wave modulation of integrated photonic resonators at microwave frequencies},}\ }\href@noop {} {\bibfield  {journal} {\bibinfo  {journal} {Nature communications}\ }\textbf {\bibinfo {volume} {5}},\ \bibinfo {pages} {5402} (\bibinfo {year} {2014})}\BibitemShut {NoStop}%
\bibitem [{\citenamefont {Zhang}\ \emph {et~al.}(2024)\citenamefont {Zhang}, \citenamefont {Cui}, \citenamefont {Chen},\ and\ \citenamefont {Fan}}]{zhang2024integrated}%
  \BibitemOpen
  \bibfield  {author} {\bibinfo {author} {\bibfnamefont {Liang}\ \bibnamefont {Zhang}}, \bibinfo {author} {\bibfnamefont {Chaohan}\ \bibnamefont {Cui}}, \bibinfo {author} {\bibfnamefont {Pao-Kang}\ \bibnamefont {Chen}}, \ and\ \bibinfo {author} {\bibfnamefont {Linran}\ \bibnamefont {Fan}},\ }\bibfield  {title} {\enquote {\bibinfo {title} {Integrated-waveguide-based acousto-optic modulation with complete optical conversion},}\ }\href@noop {} {\bibfield  {journal} {\bibinfo  {journal} {Optica}\ }\textbf {\bibinfo {volume} {11}},\ \bibinfo {pages} {184--189} (\bibinfo {year} {2024})}\BibitemShut {NoStop}%
\bibitem [{\citenamefont {Ying}\ \emph {et~al.}(2020)\citenamefont {Ying}, \citenamefont {Feng}, \citenamefont {Zhao}, \citenamefont {Dhar}, \citenamefont {Dalir}, \citenamefont {Gu}, \citenamefont {Cheng}, \citenamefont {Soref}, \citenamefont {Pan},\ and\ \citenamefont {Chen}}]{ying2020electronic}%
  \BibitemOpen
  \bibfield  {author} {\bibinfo {author} {\bibfnamefont {Zhoufeng}\ \bibnamefont {Ying}}, \bibinfo {author} {\bibfnamefont {Chenghao}\ \bibnamefont {Feng}}, \bibinfo {author} {\bibfnamefont {Zheng}\ \bibnamefont {Zhao}}, \bibinfo {author} {\bibfnamefont {Shounak}\ \bibnamefont {Dhar}}, \bibinfo {author} {\bibfnamefont {Hamed}\ \bibnamefont {Dalir}}, \bibinfo {author} {\bibfnamefont {Jiaqi}\ \bibnamefont {Gu}}, \bibinfo {author} {\bibfnamefont {Yue}\ \bibnamefont {Cheng}}, \bibinfo {author} {\bibfnamefont {Richard}\ \bibnamefont {Soref}}, \bibinfo {author} {\bibfnamefont {David~Z}\ \bibnamefont {Pan}}, \ and\ \bibinfo {author} {\bibfnamefont {Ray~T}\ \bibnamefont {Chen}},\ }\bibfield  {title} {\enquote {\bibinfo {title} {Electronic-photonic arithmetic logic unit for high-speed computing},}\ }\href@noop {} {\bibfield  {journal} {\bibinfo  {journal} {Nature communications}\ }\textbf {\bibinfo {volume} {11}},\ \bibinfo {pages} {2154} (\bibinfo {year} {2020})}\BibitemShut {NoStop}%
\bibitem [{\citenamefont {Mukherjee}\ \emph {et~al.}(2021)\citenamefont {Mukherjee}, \citenamefont {Deng}, \citenamefont {Nodjiadjim}, \citenamefont {Riet}, \citenamefont {Mismer}, \citenamefont {Guendouz}, \citenamefont {Caillaud}, \citenamefont {Bertin}, \citenamefont {Vaissiere}, \citenamefont {Luisier} \emph {et~al.}}]{mukherjee2021towards}%
  \BibitemOpen
  \bibfield  {author} {\bibinfo {author} {\bibfnamefont {Chhandak}\ \bibnamefont {Mukherjee}}, \bibinfo {author} {\bibfnamefont {Marina}\ \bibnamefont {Deng}}, \bibinfo {author} {\bibfnamefont {Virginie}\ \bibnamefont {Nodjiadjim}}, \bibinfo {author} {\bibfnamefont {Muriel}\ \bibnamefont {Riet}}, \bibinfo {author} {\bibfnamefont {Colin}\ \bibnamefont {Mismer}}, \bibinfo {author} {\bibfnamefont {Djeber}\ \bibnamefont {Guendouz}}, \bibinfo {author} {\bibfnamefont {Christophe}\ \bibnamefont {Caillaud}}, \bibinfo {author} {\bibfnamefont {Herv{\'e}}\ \bibnamefont {Bertin}}, \bibinfo {author} {\bibfnamefont {Nicolas}\ \bibnamefont {Vaissiere}}, \bibinfo {author} {\bibfnamefont {Mathieu}\ \bibnamefont {Luisier}},  \emph {et~al.},\ }\bibfield  {title} {\enquote {\bibinfo {title} {Towards monolithic indium phosphide (inp)-based electronic photonic technologies for beyond 5g communication systems},}\ }\href@noop {} {\bibfield  {journal} {\bibinfo  {journal} {Applied Sciences}\ }\textbf {\bibinfo {volume} {11}},\
  \bibinfo {pages} {2393} (\bibinfo {year} {2021})}\BibitemShut {NoStop}%
\bibitem [{\citenamefont {Behroozpour}\ \emph {et~al.}(2016)\citenamefont {Behroozpour}, \citenamefont {Sandborn}, \citenamefont {Quack}, \citenamefont {Seok}, \citenamefont {Matsui}, \citenamefont {Wu},\ and\ \citenamefont {Boser}}]{behroozpour2016electronic}%
  \BibitemOpen
  \bibfield  {author} {\bibinfo {author} {\bibfnamefont {Behnam}\ \bibnamefont {Behroozpour}}, \bibinfo {author} {\bibfnamefont {Phillip~AM}\ \bibnamefont {Sandborn}}, \bibinfo {author} {\bibfnamefont {Niels}\ \bibnamefont {Quack}}, \bibinfo {author} {\bibfnamefont {Tae-Joon}\ \bibnamefont {Seok}}, \bibinfo {author} {\bibfnamefont {Yasuhiro}\ \bibnamefont {Matsui}}, \bibinfo {author} {\bibfnamefont {Ming~C}\ \bibnamefont {Wu}}, \ and\ \bibinfo {author} {\bibfnamefont {Bernhard~E}\ \bibnamefont {Boser}},\ }\bibfield  {title} {\enquote {\bibinfo {title} {Electronic-photonic integrated circuit for 3d microimaging},}\ }\href@noop {} {\bibfield  {journal} {\bibinfo  {journal} {IEEE Journal of Solid-State Circuits}\ }\textbf {\bibinfo {volume} {52}},\ \bibinfo {pages} {161--172} (\bibinfo {year} {2016})}\BibitemShut {NoStop}%
\bibitem [{\citenamefont {Laude}(2021)}]{laude2021principles}%
  \BibitemOpen
  \bibfield  {author} {\bibinfo {author} {\bibfnamefont {Vincent}\ \bibnamefont {Laude}},\ }\bibfield  {title} {\enquote {\bibinfo {title} {Principles and properties of phononic crystal waveguides},}\ }\href@noop {} {\bibfield  {journal} {\bibinfo  {journal} {Apl Materials}\ }\textbf {\bibinfo {volume} {9}} (\bibinfo {year} {2021})}\BibitemShut {NoStop}%
\bibitem [{\citenamefont {Wang}\ \emph {et~al.}(2024)\citenamefont {Wang}, \citenamefont {Lee},\ and\ \citenamefont {Feng}}]{wang2024perspectives}%
  \BibitemOpen
  \bibfield  {author} {\bibinfo {author} {\bibfnamefont {Yanan}\ \bibnamefont {Wang}}, \bibinfo {author} {\bibfnamefont {Jaesung}\ \bibnamefont {Lee}}, \ and\ \bibinfo {author} {\bibfnamefont {Philip X-L}\ \bibnamefont {Feng}},\ }\bibfield  {title} {\enquote {\bibinfo {title} {Perspectives on phononic waveguides for on-chip classical and quantum transduction},}\ }\href@noop {} {\bibfield  {journal} {\bibinfo  {journal} {Applied Physics Letters}\ }\textbf {\bibinfo {volume} {124}} (\bibinfo {year} {2024})}\BibitemShut {NoStop}%
\bibitem [{\citenamefont {Fu}\ \emph {et~al.}(2019)\citenamefont {Fu}, \citenamefont {Shen}, \citenamefont {Xu}, \citenamefont {Zou}, \citenamefont {Cheng}, \citenamefont {Han},\ and\ \citenamefont {Tang}}]{fu2019phononic}%
  \BibitemOpen
  \bibfield  {author} {\bibinfo {author} {\bibfnamefont {Wei}\ \bibnamefont {Fu}}, \bibinfo {author} {\bibfnamefont {Zhen}\ \bibnamefont {Shen}}, \bibinfo {author} {\bibfnamefont {Yuntao}\ \bibnamefont {Xu}}, \bibinfo {author} {\bibfnamefont {Chang-Ling}\ \bibnamefont {Zou}}, \bibinfo {author} {\bibfnamefont {Risheng}\ \bibnamefont {Cheng}}, \bibinfo {author} {\bibfnamefont {Xu}~\bibnamefont {Han}}, \ and\ \bibinfo {author} {\bibfnamefont {Hong~X}\ \bibnamefont {Tang}},\ }\bibfield  {title} {\enquote {\bibinfo {title} {Phononic integrated circuitry and spin--orbit interaction of phonons},}\ }\href@noop {} {\bibfield  {journal} {\bibinfo  {journal} {Nature communications}\ }\textbf {\bibinfo {volume} {10}},\ \bibinfo {pages} {2743} (\bibinfo {year} {2019})}\BibitemShut {NoStop}%
\bibitem [{\citenamefont {Shao}\ \emph {et~al.}(2019)\citenamefont {Shao}, \citenamefont {Yu}, \citenamefont {Maity}, \citenamefont {Sinclair}, \citenamefont {Zheng}, \citenamefont {Chia}, \citenamefont {Shams-Ansari}, \citenamefont {Wang}, \citenamefont {Zhang}, \citenamefont {Lai} \emph {et~al.}}]{shao2019microwave}%
  \BibitemOpen
  \bibfield  {author} {\bibinfo {author} {\bibfnamefont {Linbo}\ \bibnamefont {Shao}}, \bibinfo {author} {\bibfnamefont {Mengjie}\ \bibnamefont {Yu}}, \bibinfo {author} {\bibfnamefont {Smarak}\ \bibnamefont {Maity}}, \bibinfo {author} {\bibfnamefont {Neil}\ \bibnamefont {Sinclair}}, \bibinfo {author} {\bibfnamefont {Lu}~\bibnamefont {Zheng}}, \bibinfo {author} {\bibfnamefont {Cleaven}\ \bibnamefont {Chia}}, \bibinfo {author} {\bibfnamefont {Amirhassan}\ \bibnamefont {Shams-Ansari}}, \bibinfo {author} {\bibfnamefont {Cheng}\ \bibnamefont {Wang}}, \bibinfo {author} {\bibfnamefont {Mian}\ \bibnamefont {Zhang}}, \bibinfo {author} {\bibfnamefont {Keji}\ \bibnamefont {Lai}},  \emph {et~al.},\ }\bibfield  {title} {\enquote {\bibinfo {title} {Microwave-to-optical conversion using lithium niobate thin-film acoustic resonators},}\ }\href@noop {} {\bibfield  {journal} {\bibinfo  {journal} {Optica}\ }\textbf {\bibinfo {volume} {6}},\ \bibinfo {pages} {1498--1505} (\bibinfo {year} {2019})}\BibitemShut {NoStop}%
\bibitem [{\citenamefont {Shin}\ \emph {et~al.}(2013)\citenamefont {Shin}, \citenamefont {Qiu}, \citenamefont {Jarecki}, \citenamefont {Cox}, \citenamefont {Olsson}, \citenamefont {Starbuck}, \citenamefont {Wang},\ and\ \citenamefont {Rakich}}]{shin2013tailorable}%
  \BibitemOpen
  \bibfield  {author} {\bibinfo {author} {\bibfnamefont {Heedeuk}\ \bibnamefont {Shin}}, \bibinfo {author} {\bibfnamefont {Wenjun}\ \bibnamefont {Qiu}}, \bibinfo {author} {\bibfnamefont {Robert}\ \bibnamefont {Jarecki}}, \bibinfo {author} {\bibfnamefont {Jonathan~A}\ \bibnamefont {Cox}}, \bibinfo {author} {\bibfnamefont {Roy~H}\ \bibnamefont {Olsson}}, \bibinfo {author} {\bibfnamefont {Andrew}\ \bibnamefont {Starbuck}}, \bibinfo {author} {\bibfnamefont {Zheng}\ \bibnamefont {Wang}}, \ and\ \bibinfo {author} {\bibfnamefont {Peter~T}\ \bibnamefont {Rakich}},\ }\bibfield  {title} {\enquote {\bibinfo {title} {Tailorable stimulated brillouin scattering in nanoscale silicon waveguides},}\ }\href@noop {} {\bibfield  {journal} {\bibinfo  {journal} {Nature communications}\ }\textbf {\bibinfo {volume} {4}},\ \bibinfo {pages} {1--10} (\bibinfo {year} {2013})}\BibitemShut {NoStop}%
\bibitem [{\citenamefont {Xu}\ \emph {et~al.}(2022)\citenamefont {Xu}, \citenamefont {Wang}, \citenamefont {Yang}, \citenamefont {Wang}, \citenamefont {Zhang}, \citenamefont {Wang}, \citenamefont {Dong}, \citenamefont {Sun}, \citenamefont {Guo},\ and\ \citenamefont {Zou}}]{xu2022high}%
  \BibitemOpen
  \bibfield  {author} {\bibinfo {author} {\bibfnamefont {Xin-Biao}\ \bibnamefont {Xu}}, \bibinfo {author} {\bibfnamefont {Jia-Qi}\ \bibnamefont {Wang}}, \bibinfo {author} {\bibfnamefont {Yuan-Hao}\ \bibnamefont {Yang}}, \bibinfo {author} {\bibfnamefont {Weiting}\ \bibnamefont {Wang}}, \bibinfo {author} {\bibfnamefont {Yan-Lei}\ \bibnamefont {Zhang}}, \bibinfo {author} {\bibfnamefont {Bao-Zhen}\ \bibnamefont {Wang}}, \bibinfo {author} {\bibfnamefont {Chun-Hua}\ \bibnamefont {Dong}}, \bibinfo {author} {\bibfnamefont {Luyan}\ \bibnamefont {Sun}}, \bibinfo {author} {\bibfnamefont {Guang-Can}\ \bibnamefont {Guo}}, \ and\ \bibinfo {author} {\bibfnamefont {Chang-Ling}\ \bibnamefont {Zou}},\ }\bibfield  {title} {\enquote {\bibinfo {title} {High-frequency traveling-wave phononic cavity with sub-micron wavelength},}\ }\href@noop {} {\bibfield  {journal} {\bibinfo  {journal} {Applied Physics Letters}\ }\textbf {\bibinfo {volume} {120}},\ \bibinfo {pages} {163503} (\bibinfo {year} {2022})}\BibitemShut {NoStop}%
\bibitem [{\citenamefont {Wiederhecker}\ \emph {et~al.}(2019)\citenamefont {Wiederhecker}, \citenamefont {Dainese},\ and\ \citenamefont {Mayer~Alegre}}]{wiederhecker2019brillouin}%
  \BibitemOpen
  \bibfield  {author} {\bibinfo {author} {\bibfnamefont {Gustavo~S}\ \bibnamefont {Wiederhecker}}, \bibinfo {author} {\bibfnamefont {Paulo}\ \bibnamefont {Dainese}}, \ and\ \bibinfo {author} {\bibfnamefont {Thiago~P}\ \bibnamefont {Mayer~Alegre}},\ }\bibfield  {title} {\enquote {\bibinfo {title} {Brillouin optomechanics in nanophotonic structures},}\ }\href@noop {} {\bibfield  {journal} {\bibinfo  {journal} {APL photonics}\ }\textbf {\bibinfo {volume} {4}} (\bibinfo {year} {2019})}\BibitemShut {NoStop}%
\bibitem [{\citenamefont {Wolff}\ \emph {et~al.}(2015)\citenamefont {Wolff}, \citenamefont {Steel}, \citenamefont {Eggleton},\ and\ \citenamefont {Poulton}}]{wolff2015stimulated}%
  \BibitemOpen
  \bibfield  {author} {\bibinfo {author} {\bibfnamefont {Christian}\ \bibnamefont {Wolff}}, \bibinfo {author} {\bibfnamefont {Michael~J}\ \bibnamefont {Steel}}, \bibinfo {author} {\bibfnamefont {Benjamin~J}\ \bibnamefont {Eggleton}}, \ and\ \bibinfo {author} {\bibfnamefont {Christopher~G}\ \bibnamefont {Poulton}},\ }\bibfield  {title} {\enquote {\bibinfo {title} {Stimulated brillouin scattering in integrated photonic waveguides: Forces, scattering mechanisms, and coupled-mode analysis},}\ }\href@noop {} {\bibfield  {journal} {\bibinfo  {journal} {Physical Review A}\ }\textbf {\bibinfo {volume} {92}},\ \bibinfo {pages} {013836} (\bibinfo {year} {2015})}\BibitemShut {NoStop}%
\bibitem [{\citenamefont {Hashimoto}\ and\ \citenamefont {Hashimoto}(2000)}]{hashimoto2000surface}%
  \BibitemOpen
  \bibfield  {author} {\bibinfo {author} {\bibfnamefont {Ken-ya}\ \bibnamefont {Hashimoto}}\ and\ \bibinfo {author} {\bibfnamefont {Ken-Ya}\ \bibnamefont {Hashimoto}},\ }\href@noop {} {\emph {\bibinfo {title} {Surface acoustic wave devices in telecommunications}}},\ Vol.\ \bibinfo {volume} {116}\ (\bibinfo  {publisher} {Springer},\ \bibinfo {year} {2000})\BibitemShut {NoStop}%
\bibitem [{\citenamefont {Sarabalis}\ \emph {et~al.}(2021)\citenamefont {Sarabalis}, \citenamefont {Van~Laer}, \citenamefont {Patel}, \citenamefont {Dahmani}, \citenamefont {Jiang}, \citenamefont {Mayor},\ and\ \citenamefont {Safavi-Naeini}}]{sarabalis2021acousto}%
  \BibitemOpen
  \bibfield  {author} {\bibinfo {author} {\bibfnamefont {Christopher~J}\ \bibnamefont {Sarabalis}}, \bibinfo {author} {\bibfnamefont {Rapha{\"e}l}\ \bibnamefont {Van~Laer}}, \bibinfo {author} {\bibfnamefont {Rishi~N}\ \bibnamefont {Patel}}, \bibinfo {author} {\bibfnamefont {Yanni~D}\ \bibnamefont {Dahmani}}, \bibinfo {author} {\bibfnamefont {Wentao}\ \bibnamefont {Jiang}}, \bibinfo {author} {\bibfnamefont {Felix~M}\ \bibnamefont {Mayor}}, \ and\ \bibinfo {author} {\bibfnamefont {Amir~H}\ \bibnamefont {Safavi-Naeini}},\ }\bibfield  {title} {\enquote {\bibinfo {title} {Acousto-optic modulation of a wavelength-scale waveguide},}\ }\href@noop {} {\bibfield  {journal} {\bibinfo  {journal} {Optica}\ }\textbf {\bibinfo {volume} {8}},\ \bibinfo {pages} {477--483} (\bibinfo {year} {2021})}\BibitemShut {NoStop}%
\bibitem [{\citenamefont {Otterstrom}\ \emph {et~al.}(2019)\citenamefont {Otterstrom}, \citenamefont {Kittlaus}, \citenamefont {Gertler}, \citenamefont {Behunin}, \citenamefont {Lentine},\ and\ \citenamefont {Rakich}}]{otterstrom2019resonantly}%
  \BibitemOpen
  \bibfield  {author} {\bibinfo {author} {\bibfnamefont {Nils~T}\ \bibnamefont {Otterstrom}}, \bibinfo {author} {\bibfnamefont {Eric~A}\ \bibnamefont {Kittlaus}}, \bibinfo {author} {\bibfnamefont {Shai}\ \bibnamefont {Gertler}}, \bibinfo {author} {\bibfnamefont {Ryan~O}\ \bibnamefont {Behunin}}, \bibinfo {author} {\bibfnamefont {Anthony~L}\ \bibnamefont {Lentine}}, \ and\ \bibinfo {author} {\bibfnamefont {Peter~T}\ \bibnamefont {Rakich}},\ }\bibfield  {title} {\enquote {\bibinfo {title} {Resonantly enhanced nonreciprocal silicon brillouin amplifier},}\ }\href@noop {} {\bibfield  {journal} {\bibinfo  {journal} {Optica}\ }\textbf {\bibinfo {volume} {6}},\ \bibinfo {pages} {1117--1123} (\bibinfo {year} {2019})}\BibitemShut {NoStop}%
\bibitem [{\citenamefont {Powell}\ \emph {et~al.}(2022)\citenamefont {Powell}, \citenamefont {Li}, \citenamefont {Shams-Ansari}, \citenamefont {Wang}, \citenamefont {Meng}, \citenamefont {Sinclair}, \citenamefont {Deng}, \citenamefont {Lon{\v{c}}ar},\ and\ \citenamefont {Yi}}]{powell2022integrated}%
  \BibitemOpen
  \bibfield  {author} {\bibinfo {author} {\bibfnamefont {Keith}\ \bibnamefont {Powell}}, \bibinfo {author} {\bibfnamefont {Liwei}\ \bibnamefont {Li}}, \bibinfo {author} {\bibfnamefont {Amirhassan}\ \bibnamefont {Shams-Ansari}}, \bibinfo {author} {\bibfnamefont {Jianfu}\ \bibnamefont {Wang}}, \bibinfo {author} {\bibfnamefont {Debin}\ \bibnamefont {Meng}}, \bibinfo {author} {\bibfnamefont {Neil}\ \bibnamefont {Sinclair}}, \bibinfo {author} {\bibfnamefont {Jiangdong}\ \bibnamefont {Deng}}, \bibinfo {author} {\bibfnamefont {Marko}\ \bibnamefont {Lon{\v{c}}ar}}, \ and\ \bibinfo {author} {\bibfnamefont {Xiaoke}\ \bibnamefont {Yi}},\ }\bibfield  {title} {\enquote {\bibinfo {title} {Integrated silicon carbide electro-optic modulator},}\ }\href@noop {} {\bibfield  {journal} {\bibinfo  {journal} {Nature Communications}\ }\textbf {\bibinfo {volume} {13}},\ \bibinfo {pages} {1851} (\bibinfo {year} {2022})}\BibitemShut {NoStop}%
\bibitem [{\citenamefont {Yamanouchi}\ and\ \citenamefont {Satoh}(2005)}]{yamanouchi2005ultra}%
  \BibitemOpen
  \bibfield  {author} {\bibinfo {author} {\bibfnamefont {Kazuhiko}\ \bibnamefont {Yamanouchi}}\ and\ \bibinfo {author} {\bibfnamefont {Yusuke}\ \bibnamefont {Satoh}},\ }\bibfield  {title} {\enquote {\bibinfo {title} {Ultra low-insertion-loss surface acoustic wave filters using unidirectional interdigital transducers with grating saw substrates},}\ }\href@noop {} {\bibfield  {journal} {\bibinfo  {journal} {Japanese journal of applied physics}\ }\textbf {\bibinfo {volume} {44}},\ \bibinfo {pages} {4532} (\bibinfo {year} {2005})}\BibitemShut {NoStop}%
\bibitem [{\citenamefont {Fall}\ \emph {et~al.}(2017)\citenamefont {Fall}, \citenamefont {Duquennoy}, \citenamefont {Ouaftouh}, \citenamefont {Smagin}, \citenamefont {Piwakowski},\ and\ \citenamefont {Jenot}}]{fall2017generation}%
  \BibitemOpen
  \bibfield  {author} {\bibinfo {author} {\bibfnamefont {Dame}\ \bibnamefont {Fall}}, \bibinfo {author} {\bibfnamefont {Marc}\ \bibnamefont {Duquennoy}}, \bibinfo {author} {\bibfnamefont {Mohammadi}\ \bibnamefont {Ouaftouh}}, \bibinfo {author} {\bibfnamefont {Nikolay}\ \bibnamefont {Smagin}}, \bibinfo {author} {\bibfnamefont {Bogdan}\ \bibnamefont {Piwakowski}}, \ and\ \bibinfo {author} {\bibfnamefont {Frederic}\ \bibnamefont {Jenot}},\ }\bibfield  {title} {\enquote {\bibinfo {title} {Generation of broadband surface acoustic waves using a dual temporal-spatial chirp method},}\ }\href@noop {} {\bibfield  {journal} {\bibinfo  {journal} {The Journal of the Acoustical Society of America}\ }\textbf {\bibinfo {volume} {142}},\ \bibinfo {pages} {EL108--EL112} (\bibinfo {year} {2017})}\BibitemShut {NoStop}%
\bibitem [{\citenamefont {Mayor}\ \emph {et~al.}(2021)\citenamefont {Mayor}, \citenamefont {Jiang}, \citenamefont {Sarabalis}, \citenamefont {McKenna}, \citenamefont {Witmer},\ and\ \citenamefont {Safavi-Naeini}}]{mayor2021gigahertz}%
  \BibitemOpen
  \bibfield  {author} {\bibinfo {author} {\bibfnamefont {Felix~M}\ \bibnamefont {Mayor}}, \bibinfo {author} {\bibfnamefont {Wentao}\ \bibnamefont {Jiang}}, \bibinfo {author} {\bibfnamefont {Christopher~J}\ \bibnamefont {Sarabalis}}, \bibinfo {author} {\bibfnamefont {Timothy~P}\ \bibnamefont {McKenna}}, \bibinfo {author} {\bibfnamefont {Jeremy~D}\ \bibnamefont {Witmer}}, \ and\ \bibinfo {author} {\bibfnamefont {Amir~H}\ \bibnamefont {Safavi-Naeini}},\ }\bibfield  {title} {\enquote {\bibinfo {title} {Gigahertz phononic integrated circuits on thin-film lithium niobate on sapphire},}\ }\href@noop {} {\bibfield  {journal} {\bibinfo  {journal} {Physical Review Applied}\ }\textbf {\bibinfo {volume} {15}},\ \bibinfo {pages} {014039} (\bibinfo {year} {2021})}\BibitemShut {NoStop}%
\bibitem [{\citenamefont {Shao}\ \emph {et~al.}(2022{\natexlab{a}})\citenamefont {Shao}, \citenamefont {Ding}, \citenamefont {Ma}, \citenamefont {Zhang}, \citenamefont {Sinclair},\ and\ \citenamefont {Lon{\v{c}}ar}}]{shao2022thermal}%
  \BibitemOpen
  \bibfield  {author} {\bibinfo {author} {\bibfnamefont {Linbo}\ \bibnamefont {Shao}}, \bibinfo {author} {\bibfnamefont {Sophie~W}\ \bibnamefont {Ding}}, \bibinfo {author} {\bibfnamefont {Yunwei}\ \bibnamefont {Ma}}, \bibinfo {author} {\bibfnamefont {Yuhao}\ \bibnamefont {Zhang}}, \bibinfo {author} {\bibfnamefont {Neil}\ \bibnamefont {Sinclair}}, \ and\ \bibinfo {author} {\bibfnamefont {Marko}\ \bibnamefont {Lon{\v{c}}ar}},\ }\bibfield  {title} {\enquote {\bibinfo {title} {Thermal modulation of gigahertz surface acoustic waves on lithium niobate},}\ }\href@noop {} {\bibfield  {journal} {\bibinfo  {journal} {Physical Review Applied}\ }\textbf {\bibinfo {volume} {18}},\ \bibinfo {pages} {054078} (\bibinfo {year} {2022}{\natexlab{a}})}\BibitemShut {NoStop}%
\bibitem [{\citenamefont {Shao}\ \emph {et~al.}(2022{\natexlab{b}})\citenamefont {Shao}, \citenamefont {Zhu}, \citenamefont {Colangelo}, \citenamefont {Lee}, \citenamefont {Sinclair}, \citenamefont {Hu}, \citenamefont {Rakich}, \citenamefont {Lai}, \citenamefont {Berggren},\ and\ \citenamefont {Lon{\v{c}}ar}}]{shao2022electrical}%
  \BibitemOpen
  \bibfield  {author} {\bibinfo {author} {\bibfnamefont {Linbo}\ \bibnamefont {Shao}}, \bibinfo {author} {\bibfnamefont {Di}~\bibnamefont {Zhu}}, \bibinfo {author} {\bibfnamefont {Marco}\ \bibnamefont {Colangelo}}, \bibinfo {author} {\bibfnamefont {Daehun}\ \bibnamefont {Lee}}, \bibinfo {author} {\bibfnamefont {Neil}\ \bibnamefont {Sinclair}}, \bibinfo {author} {\bibfnamefont {Yaowen}\ \bibnamefont {Hu}}, \bibinfo {author} {\bibfnamefont {Peter~T}\ \bibnamefont {Rakich}}, \bibinfo {author} {\bibfnamefont {Keji}\ \bibnamefont {Lai}}, \bibinfo {author} {\bibfnamefont {Karl~K}\ \bibnamefont {Berggren}}, \ and\ \bibinfo {author} {\bibfnamefont {Marko}\ \bibnamefont {Lon{\v{c}}ar}},\ }\bibfield  {title} {\enquote {\bibinfo {title} {Electrical control of surface acoustic waves},}\ }\href@noop {} {\bibfield  {journal} {\bibinfo  {journal} {Nature Electronics}\ }\textbf {\bibinfo {volume} {5}},\ \bibinfo {pages} {348--355} (\bibinfo {year} {2022}{\natexlab{b}})}\BibitemShut {NoStop}%
\bibitem [{\citenamefont {Chen}\ \emph {et~al.}(2017)\citenamefont {Chen}, \citenamefont {Fu}, \citenamefont {Huang}, \citenamefont {Zhang}, \citenamefont {Yang}, \citenamefont {Montes}, \citenamefont {Baranowski},\ and\ \citenamefont {Zhao}}]{chen2017low}%
  \BibitemOpen
  \bibfield  {author} {\bibinfo {author} {\bibfnamefont {Hong}\ \bibnamefont {Chen}}, \bibinfo {author} {\bibfnamefont {Houqiang}\ \bibnamefont {Fu}}, \bibinfo {author} {\bibfnamefont {Xuanqi}\ \bibnamefont {Huang}}, \bibinfo {author} {\bibfnamefont {Xiaodong}\ \bibnamefont {Zhang}}, \bibinfo {author} {\bibfnamefont {Tsung-Han}\ \bibnamefont {Yang}}, \bibinfo {author} {\bibfnamefont {Jossue~A}\ \bibnamefont {Montes}}, \bibinfo {author} {\bibfnamefont {Izak}\ \bibnamefont {Baranowski}}, \ and\ \bibinfo {author} {\bibfnamefont {Yuji}\ \bibnamefont {Zhao}},\ }\bibfield  {title} {\enquote {\bibinfo {title} {Low loss gan waveguides at the visible spectral wavelengths for integrated photonics applications},}\ }\href@noop {} {\bibfield  {journal} {\bibinfo  {journal} {Optics express}\ }\textbf {\bibinfo {volume} {25}},\ \bibinfo {pages} {31758--31773} (\bibinfo {year} {2017})}\BibitemShut {NoStop}%
\bibitem [{\citenamefont {Zhang}\ \emph {et~al.}(2025)\citenamefont {Zhang}, \citenamefont {Xue}, \citenamefont {Chen}, \citenamefont {Guo}, \citenamefont {Wang}, \citenamefont {Li},\ and\ \citenamefont {Yan}}]{zhang2025highly}%
  \BibitemOpen
  \bibfield  {author} {\bibinfo {author} {\bibfnamefont {Liang}\ \bibnamefont {Zhang}}, \bibinfo {author} {\bibfnamefont {Yongzhou}\ \bibnamefont {Xue}}, \bibinfo {author} {\bibfnamefont {Zewei}\ \bibnamefont {Chen}}, \bibinfo {author} {\bibfnamefont {Yanan}\ \bibnamefont {Guo}}, \bibinfo {author} {\bibfnamefont {Junxi}\ \bibnamefont {Wang}}, \bibinfo {author} {\bibfnamefont {Jinmin}\ \bibnamefont {Li}}, \ and\ \bibinfo {author} {\bibfnamefont {Jianchang}\ \bibnamefont {Yan}},\ }\bibfield  {title} {\enquote {\bibinfo {title} {Highly efficient acousto-optic modulation driven by ultra-low power in integrated photonic--phononic waveguides},}\ }\href@noop {} {\bibfield  {journal} {\bibinfo  {journal} {Laser \& Photonics Reviews}\ ,\ \bibinfo {pages} {2401952}} (\bibinfo {year} {2025})}\BibitemShut {NoStop}%
\bibitem [{\citenamefont {Ravichandran}\ and\ \citenamefont {Broido}(2020)}]{ravichandran2020phonon}%
  \BibitemOpen
  \bibfield  {author} {\bibinfo {author} {\bibfnamefont {Navaneetha~K}\ \bibnamefont {Ravichandran}}\ and\ \bibinfo {author} {\bibfnamefont {David}\ \bibnamefont {Broido}},\ }\bibfield  {title} {\enquote {\bibinfo {title} {Phonon-phonon interactions in strongly bonded solids: selection rules and higher-order processes},}\ }\href@noop {} {\bibfield  {journal} {\bibinfo  {journal} {Physical Review X}\ }\textbf {\bibinfo {volume} {10}},\ \bibinfo {pages} {021063} (\bibinfo {year} {2020})}\BibitemShut {NoStop}%
\bibitem [{\citenamefont {Chu}\ \emph {et~al.}(2017)\citenamefont {Chu}, \citenamefont {Kharel}, \citenamefont {Renninger}, \citenamefont {Burkhart}, \citenamefont {Frunzio}, \citenamefont {Rakich},\ and\ \citenamefont {Schoelkopf}}]{chu2017quantum}%
  \BibitemOpen
  \bibfield  {author} {\bibinfo {author} {\bibfnamefont {Yiwen}\ \bibnamefont {Chu}}, \bibinfo {author} {\bibfnamefont {Prashanta}\ \bibnamefont {Kharel}}, \bibinfo {author} {\bibfnamefont {William~H}\ \bibnamefont {Renninger}}, \bibinfo {author} {\bibfnamefont {Luke~D}\ \bibnamefont {Burkhart}}, \bibinfo {author} {\bibfnamefont {Luigi}\ \bibnamefont {Frunzio}}, \bibinfo {author} {\bibfnamefont {Peter~T}\ \bibnamefont {Rakich}}, \ and\ \bibinfo {author} {\bibfnamefont {Robert~J}\ \bibnamefont {Schoelkopf}},\ }\bibfield  {title} {\enquote {\bibinfo {title} {Quantum acoustics with superconducting qubits},}\ }\href@noop {} {\bibfield  {journal} {\bibinfo  {journal} {Science}\ }\textbf {\bibinfo {volume} {358}},\ \bibinfo {pages} {199--202} (\bibinfo {year} {2017})}\BibitemShut {NoStop}%
\bibitem [{\citenamefont {Manenti}\ \emph {et~al.}(2017)\citenamefont {Manenti}, \citenamefont {Kockum}, \citenamefont {Patterson}, \citenamefont {Behrle}, \citenamefont {Rahamim}, \citenamefont {Tancredi}, \citenamefont {Nori},\ and\ \citenamefont {Leek}}]{manenti2017circuit}%
  \BibitemOpen
  \bibfield  {author} {\bibinfo {author} {\bibfnamefont {Riccardo}\ \bibnamefont {Manenti}}, \bibinfo {author} {\bibfnamefont {Anton~F}\ \bibnamefont {Kockum}}, \bibinfo {author} {\bibfnamefont {Andrew}\ \bibnamefont {Patterson}}, \bibinfo {author} {\bibfnamefont {Tanja}\ \bibnamefont {Behrle}}, \bibinfo {author} {\bibfnamefont {Joseph}\ \bibnamefont {Rahamim}}, \bibinfo {author} {\bibfnamefont {Giovanna}\ \bibnamefont {Tancredi}}, \bibinfo {author} {\bibfnamefont {Franco}\ \bibnamefont {Nori}}, \ and\ \bibinfo {author} {\bibfnamefont {Peter~J}\ \bibnamefont {Leek}},\ }\bibfield  {title} {\enquote {\bibinfo {title} {Circuit quantum acoustodynamics with surface acoustic waves},}\ }\href@noop {} {\bibfield  {journal} {\bibinfo  {journal} {Nature communications}\ }\textbf {\bibinfo {volume} {8}},\ \bibinfo {pages} {975} (\bibinfo {year} {2017})}\BibitemShut {NoStop}%
\bibitem [{\citenamefont {Zheng}\ \emph {et~al.}(2021)\citenamefont {Zheng}, \citenamefont {Zhang}, \citenamefont {Song}, \citenamefont {Feng}, \citenamefont {Xu}, \citenamefont {Sun}, \citenamefont {Yang}, \citenamefont {Chen}, \citenamefont {Wei},\ and\ \citenamefont {Chen}}]{zheng2021gallium}%
  \BibitemOpen
  \bibfield  {author} {\bibinfo {author} {\bibfnamefont {Zheyang}\ \bibnamefont {Zheng}}, \bibinfo {author} {\bibfnamefont {Li}~\bibnamefont {Zhang}}, \bibinfo {author} {\bibfnamefont {Wenjie}\ \bibnamefont {Song}}, \bibinfo {author} {\bibfnamefont {Sirui}\ \bibnamefont {Feng}}, \bibinfo {author} {\bibfnamefont {Han}\ \bibnamefont {Xu}}, \bibinfo {author} {\bibfnamefont {Jiahui}\ \bibnamefont {Sun}}, \bibinfo {author} {\bibfnamefont {Song}\ \bibnamefont {Yang}}, \bibinfo {author} {\bibfnamefont {Tao}\ \bibnamefont {Chen}}, \bibinfo {author} {\bibfnamefont {Jin}\ \bibnamefont {Wei}}, \ and\ \bibinfo {author} {\bibfnamefont {Kevin~J}\ \bibnamefont {Chen}},\ }\bibfield  {title} {\enquote {\bibinfo {title} {Gallium nitride-based complementary logic integrated circuits},}\ }\href@noop {} {\bibfield  {journal} {\bibinfo  {journal} {Nature Electronics}\ }\textbf {\bibinfo {volume} {4}},\ \bibinfo {pages} {595--603} (\bibinfo {year} {2021})}\BibitemShut {NoStop}%
\bibitem [{\citenamefont {Zhang}\ \emph {et~al.}(2020{\natexlab{b}})\citenamefont {Zhang}, \citenamefont {Li}, \citenamefont {Liu}, \citenamefont {Min}, \citenamefont {Chang}, \citenamefont {Xiong}, \citenamefont {Park}, \citenamefont {Kim}, \citenamefont {Jung}, \citenamefont {Park} \emph {et~al.}}]{zhang2020heterogeneously}%
  \BibitemOpen
  \bibfield  {author} {\bibinfo {author} {\bibfnamefont {Huilong}\ \bibnamefont {Zhang}}, \bibinfo {author} {\bibfnamefont {Jinghao}\ \bibnamefont {Li}}, \bibinfo {author} {\bibfnamefont {Dong}\ \bibnamefont {Liu}}, \bibinfo {author} {\bibfnamefont {Seunghwan}\ \bibnamefont {Min}}, \bibinfo {author} {\bibfnamefont {Tzu-Hsuan}\ \bibnamefont {Chang}}, \bibinfo {author} {\bibfnamefont {Kanglin}\ \bibnamefont {Xiong}}, \bibinfo {author} {\bibfnamefont {Sung~Hyun}\ \bibnamefont {Park}}, \bibinfo {author} {\bibfnamefont {Jisoo}\ \bibnamefont {Kim}}, \bibinfo {author} {\bibfnamefont {Yei~Hwan}\ \bibnamefont {Jung}}, \bibinfo {author} {\bibfnamefont {Jeongpil}\ \bibnamefont {Park}},  \emph {et~al.},\ }\bibfield  {title} {\enquote {\bibinfo {title} {Heterogeneously integrated flexible microwave amplifiers on a cellulose nanofibril substrate},}\ }\href@noop {} {\bibfield  {journal} {\bibinfo  {journal} {Nature communications}\ }\textbf {\bibinfo {volume} {11}},\ \bibinfo {pages} {3118} (\bibinfo {year}
  {2020}{\natexlab{b}})}\BibitemShut {NoStop}%
\bibitem [{\citenamefont {Baek}\ \emph {et~al.}(2023)\citenamefont {Baek}, \citenamefont {Park}, \citenamefont {Shim}, \citenamefont {Kim}, \citenamefont {Park}, \citenamefont {Kim}, \citenamefont {Geum},\ and\ \citenamefont {Kim}}]{baek2023ultra}%
  \BibitemOpen
  \bibfield  {author} {\bibinfo {author} {\bibfnamefont {Woo~Jin}\ \bibnamefont {Baek}}, \bibinfo {author} {\bibfnamefont {Juhyuk}\ \bibnamefont {Park}}, \bibinfo {author} {\bibfnamefont {Joonsup}\ \bibnamefont {Shim}}, \bibinfo {author} {\bibfnamefont {Bong~Ho}\ \bibnamefont {Kim}}, \bibinfo {author} {\bibfnamefont {Seongchong}\ \bibnamefont {Park}}, \bibinfo {author} {\bibfnamefont {Hyun~Soo}\ \bibnamefont {Kim}}, \bibinfo {author} {\bibfnamefont {Dae-Myeong}\ \bibnamefont {Geum}}, \ and\ \bibinfo {author} {\bibfnamefont {Sang~Hyeon}\ \bibnamefont {Kim}},\ }\bibfield  {title} {\enquote {\bibinfo {title} {Ultra-low-current driven ingan blue micro light-emitting diodes for electrically efficient and self-heating relaxed microdisplay},}\ }\href@noop {} {\bibfield  {journal} {\bibinfo  {journal} {Nature Communications}\ }\textbf {\bibinfo {volume} {14}},\ \bibinfo {pages} {1386} (\bibinfo {year} {2023})}\BibitemShut {NoStop}%
\bibitem [{\citenamefont {Vafadar}\ and\ \citenamefont {Zhao}(2023)}]{vafadar2023ultralow}%
  \BibitemOpen
  \bibfield  {author} {\bibinfo {author} {\bibfnamefont {Mohammad~Fazel}\ \bibnamefont {Vafadar}}\ and\ \bibinfo {author} {\bibfnamefont {Songrui}\ \bibnamefont {Zhao}},\ }\bibfield  {title} {\enquote {\bibinfo {title} {Ultralow threshold surface emitting ultraviolet lasers with semiconductor nanowires},}\ }\href@noop {} {\bibfield  {journal} {\bibinfo  {journal} {Scientific Reports}\ }\textbf {\bibinfo {volume} {13}},\ \bibinfo {pages} {6633} (\bibinfo {year} {2023})}\BibitemShut {NoStop}%
\bibitem [{\citenamefont {Jain}\ \emph {et~al.}(2023)\citenamefont {Jain}, \citenamefont {Syed}, \citenamefont {Balendhran}, \citenamefont {Abbas}, \citenamefont {Ako}, \citenamefont {Low}, \citenamefont {Lobo}, \citenamefont {Zavabeti}, \citenamefont {Murdoch}, \citenamefont {Gupta} \emph {et~al.}}]{jain2023atomically}%
  \BibitemOpen
  \bibfield  {author} {\bibinfo {author} {\bibfnamefont {Shubhendra~Kumar}\ \bibnamefont {Jain}}, \bibinfo {author} {\bibfnamefont {Nitu}\ \bibnamefont {Syed}}, \bibinfo {author} {\bibfnamefont {Sivacarendran}\ \bibnamefont {Balendhran}}, \bibinfo {author} {\bibfnamefont {Sherif Abdulkader~Tawfik}\ \bibnamefont {Abbas}}, \bibinfo {author} {\bibfnamefont {Rajour~Tanyi}\ \bibnamefont {Ako}}, \bibinfo {author} {\bibfnamefont {Mei~Xian}\ \bibnamefont {Low}}, \bibinfo {author} {\bibfnamefont {Charlene}\ \bibnamefont {Lobo}}, \bibinfo {author} {\bibfnamefont {Ali}\ \bibnamefont {Zavabeti}}, \bibinfo {author} {\bibfnamefont {Billy~J}\ \bibnamefont {Murdoch}}, \bibinfo {author} {\bibfnamefont {Govind}\ \bibnamefont {Gupta}},  \emph {et~al.},\ }\bibfield  {title} {\enquote {\bibinfo {title} {Atomically thin gallium nitride for high-performance photodetection},}\ }\href@noop {} {\bibfield  {journal} {\bibinfo  {journal} {Advanced Optical Materials}\ ,\ \bibinfo {pages} {2300438}} (\bibinfo {year} {2023})}\BibitemShut
  {NoStop}%
\end{thebibliography}%

\vspace{4pt}
\textbf{Acknowledgments}

This work is supported by the Office of the Under Secretary of Defense for Research and Engineering under DEPSCoR program award number FA9550-21-1-0225 managed by Army Research Office.
LZ, CC, YZ, PC and LF also acknowledge the support from U.S. Department of Energy, Office of Advanced Scientific Computing Research, (Field Work Proposal ERKJ355) and Office of Naval Research (N00014-25-1-2130). LF acknowledges the support from Coherent/II-VI foundation.

\vspace{4pt}
\textbf{Author contributions}

The experiments were conceived by LF and LZ. The device was designed and fabricated by LZ. Measurements were performed by LZ, YX, and PC. Analysis of the results was conducted by LZ, LF, and CC. All authors participated in the manuscript preparation.

\vspace{4pt}
\textbf{Competing interests}

The Authors declare no Competing Financial or Non-Financial Interests

\end{document}